\documentclass[sigconf]{acmart}

\setcopyright{none}
\copyrightyear{2026}
\acmYear{2026}
\acmDOI{}
\acmISBN{}
\acmConference[Conference '26]{Conference}{June 2026}{City, Country}

\usepackage{amsmath,amsfonts,amsthm,stmaryrd}
\usepackage{algorithm}
\usepackage{algpseudocode}
\usepackage{booktabs}
\usepackage{listings}
\usepackage{xcolor}
\usepackage{tikz}
\usetikzlibrary{positioning}

\lstdefinelanguage{rust}{
  keywords={fn, let, mut, while, if, else, return, true, false, in, break, continue,
    Arc, Mutex, Condvar},
  sensitive=true,
  morecomment=[l]{//},
  morecomment=[s]{/*}{*/},
  morestring=[b]",
  morestring=[b]',
}
\usepackage{enumitem}
\usepackage{subcaption}
\usepackage{multirow}
\usepackage{url}
\usepackage{graphicx}
\usepackage{pifont}
\newcommand{\cmark}{\ding{51}}
\newcommand{\xmark}{\ding{55}}
\newtheorem{definition}{Definition}

\newtheorem{remark}{Remark}

\begin{document}

\title[CIR+CVN for Concurrent Programs]{CIR+CVN: Bridging LLM Semantic Understanding and Petri-Net Verification for Concurrent Programs}

\author{Kaiwen Zhang}
\affiliation{%
  \institution{Tongji University}
  \city{Shanghai}
  \country{China}
}
\email{zhangkw@tongji.edu.cn}

\author{Guanjun Liu}
\affiliation{%
  \institution{Tongji University}
  \city{Shanghai}
  \country{China}
}
\email{liuguanjun@tongji.edu.cn}

\begin{abstract}
Recovering concurrency structure directly from source code is difficult because shared-resource identity and protection relations are often obscured by aliasing, ownership, and API-specific idioms. We therefore study a specification-driven, model-first verification architecture for LLM-assisted concurrent program construction. Instead of verifying arbitrary source code, a large language model first synthesizes a verification-oriented concurrency artifact from a natural-language requirement or system specification. The first formalism, the Concurrency Intermediate Representation (\textsc{Cir}), is a statement-level, alias-free model in which shared resources are globally named, protection relations are explicit, and each statement carries a stable identifier. The second formalism, the Concurrency Verification Net (\textsc{Cvn}), is a weighted place/transition Petri net with a finite global store and three-valued guards for data-dependent branching. A validated \textsc{Cir} artifact is translated mechanically to \textsc{Cvn}, explored exhaustively, and any counterexample is mapped back to statement identifiers to guide targeted repair. To reduce the risk of bug-free but behavior-dropping repairs, acceptance additionally applies a lightweight goal-reachability check over designated critical outcomes. We formalize both representations, prove translation-correspondence results for deadlock and signal-loss analysis, define a two-layer checking architecture with 61 static rules and 5 analysis predicates, and evaluate the pipeline on 9 representative bounded-concurrency patterns. The results show that the method supports iterative bug detection and repair on \textsc{Cir} artifacts and that goal reachability helps filter semantically incomplete repairs. The trust boundary of the present work is the generated \textsc{Cir} artifact rather than arbitrary source code.
\end{abstract}


\maketitle

\section{Introduction}
\label{sec:introduction}
Concurrency bugs such as deadlocks, signal losses, starvation, and blocking protocol errors remain difficult to detect and repair because they arise from interactions among threads rather than from a single local control path. Dynamic tools can expose concrete failing schedules, but they do not rule out unexplored interleavings \cite{Godefroid1997VeriSoft,Musuvathi2008Heisenbugs,Burckhardt2010RandomizedScheduler}. Static techniques can, in principle, provide broader guarantees, yet they face a familiar obstacle: before reasoning about lock ordering, waiting discipline, or protocol compliance, the analysis must determine which operations refer to the same shared resource and which synchronization primitives protect which shared state \cite{Engler2003RacerX,Flanagan2000TypeBasedRace,Flanagan2003Atomicity,Boyapati2002Ownership,Naik2006StaticRaceJava}. Concurrency defects therefore remain costly in practice. They may hide through extensive testing and surface only under a narrow timing window in production, precisely because the relevant behavior is distributed across threads and schedules rather than concentrated in one obviously wrong line of code.

The difficulty is not merely that concurrent code has many interleavings; it is also that source-level reasoning must reconstruct resource identity through layers of representation detail. Consider a Rust-style program in which two threads each receive a clone of an \texttt{Arc<(Mutex<State>, Condvar)>}. To determine whether both threads operate on the same mutex--condvar pair, a verifier must trace the value through heap allocation, \texttt{Arc::clone}, destructuring, borrowing, and closure capture \cite{Boyapati2002Ownership,Pratikakis2006Locksmith,Naik2006StaticRaceJava}. Without that identity information, a deadlock detector cannot reliably build a lock-order graph, and a protection checker cannot reliably determine whether a shared-variable access is guarded by the intended lock. Existing tools cope with this barrier through familiar trade-offs: coarser analyses improve scalability but lose precision, more precise analyses often scale poorly, and dynamic tools bypass alias reconstruction by observing concrete addresses at runtime but sacrifice coverage. These trade-offs become especially painful in an automated repair workflow, where diagnostics must be both precise enough to avoid misleading the repair process and actionable enough to point to specific synchronization logic.

This paper explores a different starting point. Rather than attempting to verify arbitrary source code directly, we ask whether concurrency verification can be moved earlier in the pipeline, to the stage where an LLM constructs the program from a natural-language requirement or system specification. Source-level analysis must recover resource identity through aliasing and representation detail. In contrast, during construction, the model can be asked to state the intended shared resources and synchronization structure explicitly. The key idea of our work is therefore not that source-level alias analysis has been solved, but that the verification target can be redefined as an LLM-generated concurrency artifact in which resource identity is explicit by construction \cite{Chen2021CodeLLMs,Fan2023RepairFromLLMs,Xia2023APRLLMs}. The difficulty shifts from alias recovery to intent recovery—but intent recovery from a natural-language specification is arguably more tractable for an LLM than alias recovery from optimized source code, because the specification states resource relationships in terms the model was trained on.

This reframing leads to the first formalism in our architecture, the \emph{Concurrency Intermediate Representation} (\textsc{Cir}). \textsc{Cir} is a verification-oriented, statement-level concurrency model designed to be generated by an LLM and consumed by a formal analysis engine. Every shared resource receives a unique global name; lock-to-variable protection relations are recorded explicitly; and every statement carries a stable identifier ($\mathsf{sid}$) for diagnostics and repair. The result is not a general-purpose compiler IR, but a deliberately narrowed model of the program's concurrency skeleton: synchronization operations, thread creation and joining, and data-dependent control over shared state. Arithmetic details, string manipulation, and most I/O logic are abstracted away. This narrowing is what makes model generation tractable. The LLM does not need to emit a complete executable program; it only needs to make the concurrency architecture explicit. A deterministic checker then validates the resulting artifact against 61 rules organized into 9 categories, catching malformed or semantically inconsistent \textsc{Cir} artifacts before the more expensive verification phase.

The static checker ensures structural well-formedness but does not reason about interleavings. That task requires exhaustive state-space exploration—a capability that LLMs, by their nature, do not provide. LLMs are therefore not used as standalone verifiers. Concurrency bugs are driven by combinatorial interactions among independently plausible local actions. A three-thread circular deadlock is a useful illustration: each thread may acquire locks in a locally understandable way, yet the global system still reaches a cyclic wait. The number of relevant interleavings grows rapidly with the number of concurrent entities and synchronization operations, and exhaustive coverage over that state space is precisely where formal methods remain indispensable. In our architecture, the LLM performs the semantic heavy lifting required to recover intended resource structure and synchronization intent, while the formal engine takes over to explore the induced state space exhaustively within the bounded model. The two components are complementary rather than interchangeable: the LLM provides abstraction, and the verifier provides exhaustive reasoning over interleavings.

The choice of formal back-end is therefore central. One could use explicit-state model checking, bounded encodings, or systematic schedule exploration. Explicit-state frameworks such as SPIN are powerful, but they typically require a dedicated modeling language and an additional translation layer \cite{Holzmann1997SPIN}. Bounded encodings can handle many software-verification tasks effectively, yet synchronization constructs such as condition-variable wake-up semantics, channel blocking, and semaphore counts are not always the most natural fit for that style of encoding. Dynamic and stateless exploration techniques remain extremely valuable, but they are tied to executable artifacts and bounded schedules rather than to an abstract intermediate model \cite{Godefroid1997VeriSoft,Flanagan2005DPOR,Musuvathi2008FairStateless}. Petri nets, by contrast, offer a natural foundation for this setting \cite{Murata1989PetriNets,Esparza2008Unfoldings}. They make concurrency and resource contention structural, come with well-studied analysis techniques, and compose naturally through shared places. Standard P/T nets alone are not expressive enough for data-dependent branching, while classical colored Petri nets add considerable machinery in color domains, bindings, and inscriptions \cite{Jensen2007CPNTools}. Our second formalism, the \emph{Concurrency Verification Net} (\textsc{Cvn}), occupies a middle ground: it is a weighted P/T Petri net augmented with a finite global variable store and three-valued guards. Guards evaluate to true, false, or unknown; the third value allows the net to over-approximate branches whose conditions depend on values not tracked in the finite store, preserving soundness without requiring full data modeling. Because \textsc{Cir} has already made resource identity explicit through global naming, tokens do not need to carry identity information. This allows us to capture bounded concurrency behavior and data-dependent control without moving to the full complexity of colored-token formalisms.

\begin{table}[t]
\caption{Concurrency-verification methods.}
\label{tab:comparison}
\centering\footnotesize
\setlength{\tabcolsep}{2.6pt}
\begin{tabular}{@{}llllll@{}}
\toprule
& \begin{tabular}[c]{@{}l@{}}Model\\[-2pt] source\end{tabular}
& \begin{tabular}[c]{@{}l@{}}Alias\\[-2pt] handling\end{tabular}
& Coverage
& Liveness
& \begin{tabular}[c]{@{}l@{}}Data\\[-2pt] fidelity\end{tabular} \\
\midrule
SPIN        & manual    & in model   & exhaustive           & \cmark  & full         \\
Bounded MC  & source    & static     & $k$-bounded          & limited & bit-precise  \\
Dyn./DPOR   & executable & runtime   & partial              & \xmark  & concrete     \\
CPN Tools   & manual    & in model   & exhaustive           & \cmark  & colored      \\
\midrule
\textsc{Cvn}& LLM+spec  & by constr. & exhaustive$^\dagger$ & \cmark  & fin.\ abstr. \\
\bottomrule
\multicolumn{6}{@{}p{\columnwidth}@{}}{\scriptsize
  $^\dagger$\,Within the bounded model induced by the finite global store.
  \emph{Model source}: who constructs the verification model.
  \emph{Alias handling}: how shared-resource identity is established.
  \emph{Data fidelity}: precision of data-value modeling.}
\end{tabular}
\end{table}

\textsc{Cir} generation and \textsc{Cvn} verification are not isolated steps; they form a generate--verify--repair loop \cite{Clarke2000CEGAR,Fan2023RepairFromLLMs,Bouzenia2025RepairAgent}. A bug discovered during \textsc{Cvn} state-space search yields a structured counterexample: it identifies the bug kind, provides a trace, records the corresponding \textsc{Cir} statement IDs, and summarizes the relevant end-state information such as held resources and waiting threads. Because the feedback is anchored to $\mathsf{sid}$ values rather than vague natural-language descriptions, the LLM can revise specific synchronization logic instead of regenerating the entire artifact. The repaired \textsc{Cir} is then re-validated through the same pipeline. Acceptance is not based on bug-freedom alone. A repair is accepted only if no concurrency violation is found and the user-stated business goals remain reachable. Otherwise, an apparently safe repair may still be wrong because it eliminates the bug by silently deleting required behavior. For example, removing all \texttt{notify} calls trivially eliminates signal-loss bugs, but it also prevents any thread from ever being woken.

The trust boundary of the present paper is therefore the generated \textsc{Cir} artifact, not arbitrary source code. We verify the model that the LLM produces; we do not prove general source-to-\textsc{Cir} equivalence. That boundary is a limitation, but it is also what makes the approach analyzable. By separating semantic model construction from exhaustive state exploration, the architecture creates a tractable point of contact between LLM-assisted intent recovery and formal reasoning about interleavings. The current prototype is evaluated on Rust examples, but the core \textsc{Cir}/\textsc{Cvn} abstractions are defined over synchronization semantics rather than Rust-specific syntax. The practical motivation of the work is to obtain a verification problem that is easier to analyze by first asking the model to state explicitly what the program is trying to share and how it is trying to synchronize.

The contributions of this paper are as follows:
\begin{enumerate}[leftmargin=*]

\item A model-first verification paradigm for LLM-assisted
  concurrent-program construction that shifts shared-resource
  identification from source-level alias recovery to explicit
  declaration in a generated artifact, realized as a
  generate--verify--repair loop in which a lightweight
  goal-reachability filter reduces the risk of semantically
  incomplete repairs.

\item \textsc{Cir}, a verification-oriented concurrency
  intermediate representation with global resource naming,
  explicit protection mappings, and stable statement identifiers
  ($\mathsf{sid}$), validated by a deterministic checker that
  enforces 61 structural and semantic rules organized into 9
  categories.

\item \textsc{Cvn} is a weighted place/transition Petri net augmented with a finite global store and three-valued guards. It is designed to bridge the gap between standard P/T nets, which lack support for data-dependent branching, and colored Petri nets, which are unnecessary when resource identity is already explicit.

\item A mechanical translation from \textsc{Cir} to
  \textsc{Cvn}, together with simulation-style correspondence
  results showing that deadlock and signal-loss verdicts on the
  net faithfully reflect properties of the \textsc{Cir} artifact.

\item An experimental evaluation on 9 representative
  bounded-concurrency patterns demonstrating iterative bug
  detection and repair, and showing that goal-reachability checks
  filter semantically incomplete repairs that would otherwise pass safety verification.

\end{enumerate}

Section~\ref{sec:motivation} presents the motivating example and the alias-reframing intuition.
Section~\ref{sec:architecture} presents the system architecture and repair hierarchy.
Section~\ref{sec:formal-defs} defines \textsc{Cir}, \textsc{Cvn}, and the translation between them.
Section~\ref{sec:analysis-repair} describes bug detection, structured counterexamples, and repair.
Section~\ref{sec:properties} states the formal properties of the translation and analysis.
Section~\ref{sec:evaluation} presents the experimental evaluation.
Section~\ref{sec:related} surveys related work.
Section~\ref{sec:conclusion} concludes.

\section{Motivation}
\label{sec:motivation}
A condition-variable example introduces the problem setting and illustrates how the pipeline detects bugs and guides repair.

\subsection{A Signal-Loss Example}
\label{subsec:running-example}

Figure~\ref{fig:example-a} shows a Rust program with a signal-loss bug. The worker calls \texttt{wait} without a surrounding loop; the notifier calls \texttt{notify\_one} before setting the flag. Both operations reside in the same critical section. If the notifier acquires the lock first, the signal fires with no waiter present and is lost; the worker then enters \texttt{wait} and blocks indefinitely because no second notification will arrive.

\begin{figure}[t]
\begin{lstlisting}[language=rust]
fn worker(shared: Arc<(Mutex<bool>, Condvar)>) {
    let (m, cv) = &*shared;
    let g = m.lock().unwrap();
    let _g = cv.wait(g).unwrap(); // BUG: bare wait
}
fn notifier(shared: Arc<(Mutex<bool>, Condvar)>) {
    let (m, cv) = &*shared;
    let mut g = m.lock().unwrap();
    cv.notify_one();   // notify before write
    *g = true;
}
\end{lstlisting}
\caption{Signal-loss bug: bare wait without a loop guard.}
\label{fig:example-a}
\end{figure}

Figure~\ref{fig:cir-a} shows the corresponding \textsc{Cir} artifact---not extracted from source code, but produced by an LLM from a natural-language specification. Alias analysis is unnecessary: \texttt{m0}, \texttt{cv0}, and \texttt{ready} are unique global names; the declaration \texttt{protection:\ ready:\ [m0]} makes the guarding relation explicit; and every statement carries a stable $\mathsf{sid}$ that anchors diagnostics across repair iterations.

\begin{figure}[t]
\begin{lstlisting}[basicstyle=\small\ttfamily, frame=single]
resources:
  m0:    { kind: Mutex }
  cv0:   { kind: Condvar, paired_with: m0 }
  ready: { kind: Var, type: Bool, init: false }
protection:  ready: [m0]
threads:
  worker:
    body:
    - { sid: w1, op: lock(m0),        next: w2 }
    - { sid: w2, op: wait(cv0, m0),   next: w3 }
    - { sid: w3, op: unlock(m0) }
  notifier:
    body:
    - { sid: n1, op: lock(m0),            next: n2 }
    - { sid: n2, op: notify_one(cv0),     next: n3 }
    - { sid: n3, op: write(ready, true),  next: n4 }
    - { sid: n4, op: unlock(m0) }
\end{lstlisting}
\caption{\textsc{Cir} for the signal-loss example.}
\label{fig:cir-a}
\end{figure}

\subsection{Bug Detection and Repair Guidance}
\label{subsec:repair-demo}

The \textsc{Cir} artifact is translated mechanically into a \textsc{Cvn} and explored exhaustively. The verifier detects the signal-loss bug and emits the structured repair report shown in Figure~\ref{fig:repair-report}. The report identifies the bug kind, anchors the root cause to specific $\mathsf{sid}$ values, summarizes the end-state, and provides a concrete repair suggestion---all expressed in terms of \textsc{Cir} constructs rather than Petri-net internals.

\begin{figure}[t]
\begin{lstlisting}[basicstyle=\small\ttfamily, frame=single]
bug_kind:  signal_loss
blame:     [n2, w2]
end_state:
  V[cv0_worker]: waiting
  V[ready]:      true
  held:          {}
  waiting:       [worker]
trace:     [n1, n2(lost), n3, n4, w1, w2(blocked)]
diagnosis:
  notify_one at sid=n2 fires when no thread is
  waiting on cv0; worker enters bare wait at
  sid=w2 with no loop guard and no future
  notification.
repair_suggestion:
  Insert read(ready) with branch before w2;
  add back-edge from post-wakeup to the condition
  check, forming a while(!ready) loop around
  wait(cv0, m0).
goal_check:
  critical_outcome: "worker exits synchronization"
  status: UNREACHABLE in current artifact
  note: bare wait at w2 depends on notification
        that may never arrive.
\end{lstlisting}
\caption{Structured repair report for the signal-loss example.}
\label{fig:repair-report}
\end{figure}

Three aspects of this report merit attention. First, the \emph{blame anchors} $\mathsf{sid}{=}n2$ and $\mathsf{sid}{=}w2$ tell the LLM exactly which statements to modify, avoiding regeneration of the entire artifact. Second, the \emph{repair suggestion} is expressed as a structural \textsc{Cir} edit (insert a read-and-branch node, add a back-edge), which the LLM can apply directly. Third, the \emph{goal check} verifies that the designated critical outcome is reachable after repair; a fix that achieves safety by deleting behavior (e.g., removing \texttt{notify\_one}) is rejected as behaviorally incomplete.

The same pipeline handles other anti-patterns with different diagnostics. For instance, if the worker uses \texttt{if} instead of \texttt{while} but the notifier uses the correct write-before-notify order, the verifier reports a \emph{spurious-wakeup violation}: a
non-deterministic wakeup causes the worker to exit the \texttt{if} block without re-checking the predicate. The blame anchors shift to the \texttt{if} branch and the post-wakeup point, and the repair suggestion becomes replace \texttt{if} with \texttt{while} by adding a back-edge. The goal check confirms that the worker can still reach the post-synchronization code with the predicate satisfied.

\medskip
\noindent
In summary, \textsc{Cir} eliminates alias reconstruction through global naming; \textsc{Cvn} detects bugs through exhaustive exploration; and $\mathsf{sid}$-anchored reports with goal-reachability checks guide targeted, semantically complete repairs.

\section{System Architecture}
\label{sec:architecture}

\begin{figure}[t]
\centering
\begin{tikzpicture}[
  box/.style={rectangle, draw, rounded corners=2pt, thick,
    text width=3.4cm, minimum height=0.7cm, align=center,
    font=\footnotesize},
  acc/.style={box, fill=black!8},
  node distance=0.8cm
]

\node[box]               (s1) {Stage 1: \textsc{Cir} Generation};
\node[box, below=of s1]  (s2) {Stage 2: Static Checking};
\node[box, below=of s2]  (s3) {Stage 3: \textsc{Cir}$\;\to\;$\textsc{Cvn}};
\node[box, below=of s3]  (s4) {Stage 4: State-Space Analysis};
\node[box, below=of s4]  (s5) {Stage 5: Goal Reachability};
\node[acc, below=of s5]  (ok) {Accepted};

\draw[->] (s1.south) -- (s2.north);
\draw[->] (s2.south) -- (s3.north)
  node[midway, right, font=\scriptsize] {pass};
\draw[->] (s3.south) -- (s4.north);
\draw[->] (s4.south) -- (s5.north)
  node[midway, right, font=\scriptsize] {no bug};
\draw[->] (s5.south) -- (ok.north)
  node[midway, right, font=\scriptsize] {reachable};

\draw[->, dashed] ([yshift=2pt]s2.east)
  .. controls +(0.7,0.35) and +(0.7,-0.35) ..
  ([yshift=-2pt]s2.east);
\node[font=\scriptsize] at ([xshift=10mm]s2.east) {Tier 1};

\draw[->, dashed] (s2.west) -- ++(-6mm,0) |- (s1.west);
\node[font=\scriptsize, anchor=east]
  at ([xshift=-6mm, yshift=4.5mm]s2.west) {Tier 2};

\draw[->, dashed] (s4.east) -- ++(11mm,0)
  |- ([yshift=-1mm]s1.east);
\node[font=\scriptsize, anchor=west]
  at ([xshift=11mm, yshift=5mm]s4.east) {Tier 3};

\draw[->, dashed] (s5.east) -- ++(16mm,0)
  |- ([yshift=1mm]s1.east);
\node[font=\scriptsize, anchor=west]
  at ([xshift=16mm, yshift=5mm]s5.east) {goal fail};

\end{tikzpicture}
\caption{The generate--verify--repair pipeline.
  Solid arrows show the main flow; dashed arrows show
  repair feedback loops.}
\Description{A five-stage vertical pipeline.  Stage~1 generates
  a CIR artifact via LLM.  Stage~2 performs static checking
  (Layer~1) with a Tier-1 auto-fix self-loop and a Tier-2
  feedback path to Stage~1.  Stage~3 translates CIR to CVN.
  Stage~4 performs state-space analysis (Layer~2) with a Tier-3
  feedback path to Stage~1.  Stage~5 checks goal reachability
  with a failure feedback path to Stage~1.  The artifact is
  accepted only after passing all stages.}
\label{fig:pipeline}
\end{figure}
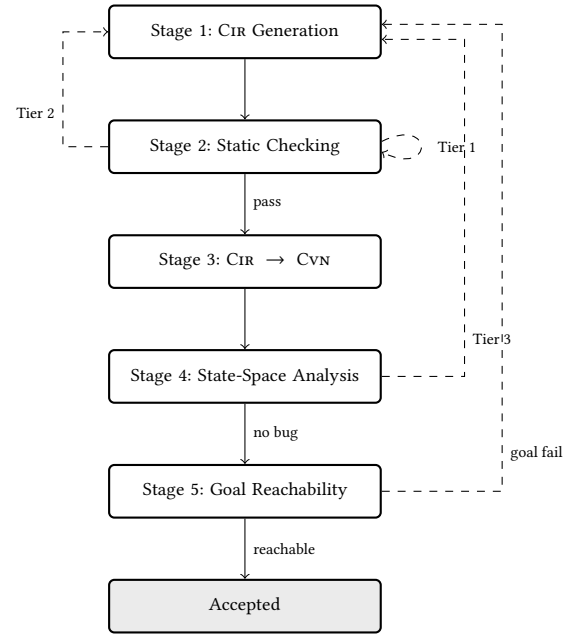

The architecture organizes five stages into two verification layers and three repair tiers.
Layer~1 (Stage~2) validates model quality through deterministic static checking; Layer~2 (Stage~4) analyzes concurrency behavior through exhaustive state-space search; and a dedicated
goal-reachability check (Stage~5) guards against behavior-dropping repairs. Figure~\ref{fig:pipeline} shows the complete pipeline.

\subsection{Pipeline Stages}
\label{subsec:pipeline-overview}

\emph{Stage~1: \textsc{Cir} generation.}
An LLM reads a natural-language requirement or system specification and produces a \textsc{Cir} artifact. The LLM is asked to recover the intended shared resources, protection mappings, and statement-level control flow rather than to emit an executable program.

\emph{Stage~2: Static checking (Layer~1).}
A deterministic checker validates the \textsc{Cir} against 61~well-formedness rules organized in 9~categories. This stage catches malformed artifacts early: missing fields, undefined resources, incompatible operations, broken spawn/join structure, and protection-mapping errors.

\emph{Stage~3: Translation to \textsc{Cvn}.}
A stateless translator converts the validated \textsc{Cir} into a \textsc{Cvn}. Each \textsc{Cir} statement gives rise to at least one anchored transition; no optimization or transition merging is performed. This keeps the correctness argument simple and preserves diagnostic traceability.

\emph{Stage~4: State-space analysis (Layer~2).}
The \textsc{Cvn} is explored through exhaustive reachability and SCC-based liveness checks. This stage runs entirely off-prompt and does not consume LLM budget.

\emph{Stage~5: Goal-reachability check.}
If Stage~4 finds no concurrency violation, the verifier checks whether each user-designated critical outcome remains reachable in the explored state space. An artifact is accepted only if it is both violation-free and goal-reachable. An apparently safe repair that achieves bug-freedom by deleting required behavior is rejected at this stage. The loop iterates until the artifact passes all stages, or until a configurable iteration budget (default: 5~rounds) is exhausted, in which case the pipeline reports failure.

\subsection{Verification and Repair Hierarchy}
\label{subsec:repair-tier} 
The two verification layers separate model quality from concurrency behavior. Three repair tiers determine the appropriate response.

\emph{Tier~1: deterministic auto-fix.}
When Layer~1 detects a non-semantic issue: a missing \texttt{unlock} at function exit, a duplicate $\mathsf{sid}$, an omitted \texttt{transfer} field, or a similar mechanical defect, the checker applies a pre-defined rewrite directly without involving the LLM.

\emph{Tier~2: structural re-generation.}
When Layer~1 detects a semantic structural problem such as an undefined resource, a broken spawn/join topology, or an inconsistent protection mapping, the diagnostic is sent to the LLM for localized regeneration of the affected \textsc{Cir} fragment.

\emph{Tier~3: logic-driven repair.}
When Layer~2 finds a concurrency violation, the verifier constructs a structured diagnostic from the bug witness including blame anchors, end-state summary, and repair suggestion and asks the
LLM to revise the affected synchronization logic. This is the most expensive repair tier and carries the main methodological novelty.

\begin{table}[t]
\caption{Verification and repair hierarchy.}
\label{tab:two-layer}
\centering\footnotesize
\begin{tabular}{@{}p{2.8cm}ll@{}}
\toprule
\textbf{Issue} & \textbf{Layer} & \textbf{Tier} \\
\midrule
\multicolumn{3}{@{}l}{\textit{Structural (Layer~1)}} \\
\quad Missing unlock/drop       & \textsc{Cir} & 1 (auto-fix) \\
\quad Duplicate $\mathsf{sid}$  & \textsc{Cir} & 1 (auto-fix) \\
\quad Undefined resource        & \textsc{Cir} & 2 (LLM) \\
\quad Spawn/join mismatch       & \textsc{Cir} & 2 (LLM) \\
\quad Protection inconsistency  & \textsc{Cir} & 2 (LLM) \\
\midrule
\multicolumn{3}{@{}l}{\textit{Behavioral (Layer~2)}} \\
\quad Deadlock                  & \textsc{Cvn} & 3 (LLM) \\
\quad Signal loss               & \textsc{Cvn} & 3 (LLM) \\
\quad Starvation                & \textsc{Cvn} & 3 (LLM) \\
\quad Cross-primitive deadlock  & \textsc{Cvn} & 3 (LLM) \\
\bottomrule
\end{tabular}
\end{table}

This hierarchy serves two purposes. First, it prevents expensive Layer~2 analysis from being wasted on malformed artifacts that should have been rejected earlier. Second, it ensures that the LLM receives feedback at the appropriate level of abstraction: mechanical corrections are applied silently, structural feedback is provided when the model is malformed, execution-grounded feedback is provided when the concurrency logic is wrong, and goal-violation feedback is provided
only after bug-freedom has been established.

\section{Formal Definitions}
\label{sec:formal-defs}

Three core formalisms underlie the method: \textsc{Cir}, \textsc{Cvn}, and the translation from \textsc{Cir} to \textsc{Cvn}.

\subsection{CIR: Concurrency Intermediate Representation}
\label{subsec:cir-def}

\begin{definition}[Function and statement]
\label{def:statement}
A \emph{function definition} is a triple $(f,\,k,\,B)$ where $f$ is the name, $k\in\{\textsf{normal},\textsf{async},\textsf{closure}\}$ is the kind, and $B=(s_1,\dots,s_n)$ is a body of statements.
A \emph{statement} is a triple $s=(\mathit{sid},\;\mathit{op},\;\mathit{transfer})$, where $\mathit{sid}\in\mathit{SID}$ is a unique identifier, $\mathit{op}$ is an operation from Table~\ref{tab:cir-ops} or the distinguished no-operation $\textsf{nop}$, and $\mathit{transfer}$ specifies the control-flow successor:
   \[
      \mathit{transfer} \in
      \bigl\{\,
        \textsf{next}(s'),\;
        \textsf{branch}(\beta,s_t,s_f),\;
        \textsf{switch}(x,\langle v_i\!\to\!s_i\rangle_{i},s_d),\;
        \textsf{return}
      \,\bigr\}
    \]
    where $\beta\in\mathit{BoolExpr}$ (Definition~\ref{def:expressions}). The operation and transfer are orthogonal: $\mathit{op}$ specifies the resource or concurrency action; $\mathit{transfer}$ specifies the successor.
\end{definition}

$\mathit{SID}$ denotes the universe of statement identifiers. Every function $(f,k,B)$ has a distinguished \emph{entry identifier} $\mathit{sid}^{0}_{f}$ (the $\mathit{sid}$ of the first statement in $B$) and a distinguished \emph{return identifier} $\mathit{ret}_{f}\notin\{\mathit{sid}\mid(\mathit{sid},\_,\_)\in B\}$. When unambiguous we write $\mathit{sid}^{0}$ and $\mathit{ret}$ without the function subscript. The sentinel $\mathit{sid}_{\bot}$ is reserved for synthetic transitions generated from function summaries.

\begin{definition}[\textsc{Cir} artifact]
\label{def:cir}
A \textsc{Cir} artifact is a 6-tuple
$\mathcal{I}=(R,\;G,\;F,\;\Sigma,\;e,\;\Gamma)$, where $R$ is a finite resource table mapping names to kinds and configurations, $G:\mathit{VarName}\rightharpoonup\mathcal{P}(\mathit{LockName})$ is a protection mapping, $F$ is a finite set of function definitions $(f,k,B)$, $\Sigma$ is a partial function-summary map for external calls, $e\in F$ is the designated entry function, and $\Gamma$ is a finite set of \textsc{Cir} business goals in Definition~\ref{def:cir-goal}.
\end{definition}

Every resource in $R$ has a unique global name. Pointers, heap indirection, and aliasing are strictly forbidden; functions reference resources directly by name. The protection mapping~$G$ is used exclusively for static checking and does not produce \textsc{Cvn} structures.

\begin{table}[t]
\caption{\textsc{Cir} operations.}
\label{tab:cir-ops}
\centering\footnotesize
\begin{tabular}{@{}llp{3.2cm}@{}}
\toprule
\textbf{Category} & \textbf{Operation} & \textbf{Target} \\
\midrule
\multirow{2}{*}{Lock}
  & \textsf{lock}, \textsf{drop}
  & Mutex, RwLock \\
  & \textsf{read\_lock}, \textsf{write\_lock}
  & RwLock \\[3pt]
\multirow{3}{*}{Synchronization}
  & \textsf{wait}, \textsf{notify\_one}, \textsf{notify\_all}
  & Condvar \\
  & \textsf{acquire}, \textsf{release}
  & Semaphore \\
  & \textsf{send}, \textsf{recv}
  & Channel \\[3pt]
\multirow{3}{*}{Data}
  & \textsf{read}, \textsf{write}
  & Var \\
  & \textsf{load}, \textsf{store}
  & Atomic \\
  & $\textsf{cas}(\mathit{expected},\mathit{new})$
  & Atomic \\[3pt]
\multirow{3}{*}{Control}
  & \textsf{spawn}, \textsf{join}
  & function (OS thread) \\
  & \textsf{spawn\_async}, \textsf{await}
  & function (async task) \\
  & \textsf{call}
  & function \\
\bottomrule
\end{tabular}
\end{table}

The summary map~$\Sigma$ provides a summary for each function whose body is not modeled in~$F$. A summary is a tuple $\Sigma(f)=(\mathit{reads},\mathit{writes},\mathit{calls},\mathit{has\_concurrency})$, where $\mathit{reads}$ and $\mathit{writes}$ are sets of resource names, $\mathit{calls}$ is a set of callee names, and $\mathit{has\_concurrency}$ is a boolean. Only the $\mathit{writes}$ field affects the \textsc{Cvn} translation; the remaining fields serve the static checker. The static checker enforces 61 well-formedness rules organized in 9~categories (codes \texttt{E0xx}--\texttt{E8xx}) covering structural integrity, name resolution, type compatibility, resource pairing, lock safety, control flow, protection mapping, and function-summary consistency. The full catalogue appears in Table~\ref{tab:cir-errors} (Appendix~\ref{app:tables}).

\begin{definition}[\textsc{Cir} business goal]
  \label{def:cir-goal}
  A \textsc{Cir} business goal is a pair $\gamma=(C_\gamma,\,V_\gamma)$, where $C_\gamma$ is a set of \emph{completion} and
  \emph{availability} requirements that $(f,\textsf{completed})$ requires function~$f$ to have returned, $(r,\textsf{available})$ requires resource~$r$ to be in its initial (free) state, and $V_\gamma:\mathit{VarName}\rightharpoonup\mathit{Val}$ specifies required variable values. Goals are expressed entirely in \textsc{Cir} vocabulary and do not
  reference \textsc{Cvn} places or markings.
\end{definition}

\begin{figure}[t]
\begin{lstlisting}[basicstyle=\small\ttfamily,frame=single]
goals:
  - id: G1
    desc: "Both threads complete, lock released,
           ready flag set"
    completion:
      - [worker, completed]
      - [notifier, completed]
    availability:
      - [m0, available]
    variables:
      ready: true
  - id: G2
    desc: "Worker reaches completion state"
    completion:
      - [worker, completed]
\end{lstlisting}
\caption{Business goals expressed in \textsc{Cir} vocabulary.}
\Description{Two CIR business goals using completion and availability requirements.}
\label{fig:goals-example}
\end{figure}

Business goals capture what the program is \emph{supposed to achieve}, not merely what it should avoid. A buggy program may fail to reach its goals because a deadlock or signal loss blocks the execution path. Conversely, a repair that eliminates all concurrency bugs may still fail to reach a goal if it inadvertently removes a critical operation. The acceptance criterion for a verified \textsc{Cir} is therefore the conjunction: (1)~no concurrency bug is detected, and (2)~every business goal is reachable. Goal reachability is checked only after bug-freedom is established, because bugs themselves can block reachability. Goals can be specified in two ways. The user can state them explicitly as part of the natural-language specification, or the LLM can derive them automatically when generating the \textsc{Cir} artifact. In either case the goals become part of~$\mathcal{I}$ and travel through the entire pipeline.

\subsection{CVN: Concurrency Verification Net}
\label{subsec:cvn-def}

\begin{definition}[Value domain]
\label{def:val}
The value domain is $\mathit{Val}=\mathit{Concrete}(\tau)\cup\{\top\}$, where $\tau$ ranges over the base types $\{\textsf{Bool},\textsf{Int},\textsf{Float},\textsf{String},\textsf{Enum}\}$. The element~$\top$ (\emph{unknown}) is absorbing: any arithmetic operation receiving~$\top$ as an operand produces~$\top$; any comparison receiving~$\top$ produces the truth value $\mathsf{unknown}$. A literal write overwrites~$\top$ unconditionally.
\end{definition}

\begin{definition}[Expressions]
\label{def:expressions}
\emph{Value expressions} and \emph{boolean expressions} over the value domain are defined by:
\begin{align*}
  \mathit{Expr} &::=
    \mathit{Lit}(v) \mid
    \mathit{Ref}(x) \mid
    \mathit{BinOp}(\mathit{Expr},\oplus,\mathit{Expr}) \\[4pt]
  \mathit{BoolExpr} &::=
    \textsf{True} \mid \textsf{False} \mid
    \mathit{Cmp}(\mathit{Expr},\mathit{cmp},\mathit{Expr}) \\
    &\quad\mid\;
    \mathit{And}(\mathit{BoolExpr},\mathit{BoolExpr}) \mid
    \mathit{Or}(\mathit{BoolExpr},\mathit{BoolExpr}) 
    &\quad\mid\;
    \mathit{Not}(\mathit{BoolExpr})
\end{align*}
Given a valuation~$V$, $\mathit{eval}(e,V)$ evaluates a value expression to an element of $\mathit{Val}$; $\mathit{Ref}(x)$ evaluates to $V(x)$ and $\mathit{BinOp}$ follows the absorbing rule for~$\top$.
\end{definition}

\begin{definition}[Guard evaluation]
  \label{def:guard-eval}
  A \emph{guard} is a $\mathit{BoolExpr}$. Given a valuation~$V$, the evaluation $\mathit{eval}(g,V)\in\{\mathsf{true},\mathsf{false},\mathsf{unknown}\}$ is defined by the absorbing rules: a comparison with a~$\top$ operand evaluates to $\mathsf{unknown}$; $\mathsf{unknown}\wedge\mathsf{false}=\mathsf{false}$;\; $\mathsf{unknown}\wedge\mathsf{true}=\mathsf{unknown}$;\; $\mathsf{unknown}\vee\mathsf{true}=\mathsf{true}$;\; $\mathsf{unknown}\vee\mathsf{false}=\mathsf{unknown}$. When a guard evaluates to~$\mathsf{unknown}$ the transition is treated as enabled; for a branch, both successors are enabled   simultaneously. This over-approximation ensures no reachable state is missed at the cost of exploring potentially infeasible paths.
\end{definition}

\begin{definition}[\textsc{Cvn}]
\label{def:cvn}
A Concurrency Verification Net is an 8-tuple
\[
  \mathcal{N}=(P,\;T,\;W_{\!\mathit{in}},\;W_{\!\mathit{out}},\;
    G,\;U,\;\mathit{Var},\;\mu)
\]
where
\begin{itemize}
  \item $P=P_c\uplus P_r\uplus P_a$ is a finite set of \emph{places}, partitioned into control places~$P_c$, resource places~$P_r$, and auxiliary places~$P_a$;
  \item $T$ is a finite set of \emph{transitions};
  \item $W_{\!\mathit{in}}:P\times T\to\mathbb{N}_0$ is the \emph{input weight function}($W_{\!\mathit{in}}(p,t)=0$ means no arc from $p$ to $t$);
  \item $W_{\!\mathit{out}}:T\times P\to\mathbb{N}_0$ is the \emph{output weight function};
  \item $G:T\to\mathit{BoolExpr}$ is the \emph{guard function} (default $\textsf{True}$);
  \item $U:T\to(\mathit{Var}\rightharpoonup\mathit{Expr})$ is the \emph{update function}, assigning each transition a
        partial map from variables to expressions (default~$\emptyset$);
  \item $\mathit{Var}$ is a finite set of variable names; and
  \item $\mu:T\to\mathit{SID}$ is the \emph{anchor mapping}.
\end{itemize}
\end{definition}

\begin{remark}[Place taxonomy and initial tokens]
\label{rem:places}
The three place classes and their initial token counts are:
\begin{enumerate}
  \item \textbf{Control places} $P_c$. One place $\mathit{cp}(f,\mathit{sid})$ per statement per function, plus a return place $\mathit{cp}(f,\mathit{ret})$. A token in $\mathit{cp}(f,\mathit{sid})$ means a thread is at statement $\mathit{sid}$ in~$f$. Initially, each spawned function's entry place $\mathit{cp}(f,\mathit{sid}^{0}_{f})$ holds one token; all other control places hold zero.

  \item \textbf{Resource places} $P_r$. One place $\mathit{rp}(r)$ per synchronization primitive. Initial token counts:
    Mutex~$\mapsto1$;\; RwLock~$\mapsto N$ where $N$ is the total number of concurrent entities;\; Semaphore$(n)\mapsto n$;\;
    Channel~$\mapsto 0$;\; Condvar signal place~$\mapsto 0$. \textsf{Var} and \textsf{Atomic} resources have no resource
    places; their state resides in the valuation~$V$.

  \item \textbf{Auxiliary places} $P_a$. For each \textsf{wait} site~$\mathit{sid}$ on a condition variable: a \emph{wait place}~$\mathit{wp}(\mathit{sid})$ (token = thread blocked) and a \emph{reacquire place}~$\mathit{ra}(\mathit{sid})$ (token = thread woken, awaiting mutex). All auxiliary places start at zero.
\end{enumerate}
\end{remark}


\begin{remark}[Anchor mapping]
\label{rem:anchor}
$\mu$ is total on transitions generated from \textsc{Cir} statements. For compound operations such as \textsf{wait}, which generates multiple transitions (WaitEnter, Wake1, WakeA, Reacquire), each transition is anchored to the originating \textsf{wait} statement's $\mathit{sid}$. Transitions synthesized from summaries are anchored to $\mathit{sid}_{\bot}$ and excluded from witness traces.
\end{remark}

\begin{definition}[\textsc{Cvn} state]
\label{def:cvn-state}
A \textsc{Cvn} state is a pair $S=(M,V)$ where $M:P\to\mathbb{N}_0$ is a marking and $V:\mathit{Var}\to\mathit{Val}$ is a valuation. The initial state $S_0=(I_m,I_v)$ assigns token counts per Remark~\ref{rem:places} and initial values per the \textsc{Cir} resource table~$R$. The pair $(\mathcal{N},S_0)$ fully determines the reachable state space.
\end{definition}

\begin{definition}[Enabling and firing]
\label{def:enabling-firing}
A transition $t\in T$ is \emph{enabled} at state $S=(M,V)$ iff
\begin{enumerate}
  \item $\forall p\in P:\; M(p)\geq W_{\!\mathit{in}}(p,t)$, and
  \item $\mathit{eval}(G(t),V)\neq\mathsf{false}$.
\end{enumerate}
Upon firing, the successor state $S'=(M',V')$ is:
\[
  M'(p)=M(p)-W_{\!\mathit{in}}(p,t)+W_{\!\mathit{out}}(t,p)
  \quad\text{for each }p\in P,
\]
\[
  V'(x)=
  \begin{cases}
    \mathit{eval}\bigl(U(t)(x),\;V\bigr)
      & \text{if } x\in\mathrm{dom}(U(t)),\\
    V(x) & \text{otherwise.}
  \end{cases}
\]
All expressions in $U(t)$ are evaluated with respect to the pre-firing valuation~$V$.
If multiple transitions are simultaneously enabled, the search explores all resulting interleavings.
\end{definition}

Each transition also carries a classification tag (Lock, Unlock, Spawn, etc.) for diagnostics and visualization; the tag does not affect firing semantics.

\subsection{CIR to CVN Translation}
\label{subsec:translation}

Let $\textsc{Cir}_v$ denote the subset of \textsc{Cir} artifacts that pass all 61 static-checking rules. The translator is a pure, total function $\mathit{translate}:\textsc{Cir}_v\to\textsc{Cvn}$
executed in three phases.

\paragraph{Phase 1: Resource scanning.}
For each synchronization primitive in~$R$, generate a resource place $\mathit{rp}(r)$ and set its initial token count per Remark~\ref{rem:places}. Compute the RwLock concurrency bound~$N$ as the total number of concurrent entities. For each \textsf{Var} and \textsf{Atomic} resource, initialize the corresponding entry in the valuation~$I_v$. For each \textsf{Condvar}, generate the auxiliary places and variables described in Remarks~\ref{rem:places} and~\ref{rem:condvar-vars}.

\paragraph{Phase 2: Function-body translation.}
For each function in~$F$ and each statement in its body, generate control places, transitions, weights, guards, and updates
according to Table~\ref{tab:translation-rules} (Appendix~\ref{app:tables}).

\paragraph{Phase 3: Summary translation.}
For each function referenced by a \textsf{call} that has a summary in~$\Sigma$ but no body in~$F$, generate a single atomic
transition. For each resource~$x$ in the $\mathit{writes}$ set, the update sets $x:=\top$.


$\mathit{translate}$ also maps each \textsc{Cir} goal $\gamma=(C_\gamma,V_\gamma)$ to a \textsc{Cvn} reachability query: $(f,\textsf{completed})$ maps to $M(\mathit{cp}(f,\mathit{ret}))\geq 1$; $(r,\textsf{available})$ maps to $M(\mathit{rp}(r))\geq I_m(\mathit{rp}(r))$; variable requirements pass through unchanged. The mapping is purely internal: users specify and receive results in \textsc{Cir} vocabulary.

\begin{definition}[Goal satisfaction]
\label{def:goal-sat}
A \textsc{Cvn} state $(M,V)$ \emph{satisfies} a \textsc{Cir} goal $\gamma=(C_\gamma,V_\gamma)$, written $(M,V)\models\gamma$, iff every completion and availability requirement holds in~$M$ and $\forall x\in\mathrm{dom}(V_\gamma):\;V(x)=V_\gamma(x)$.
A goal is \emph{reachable} if at least one reachable state satisfies it. The existential criterion is appropriate because concurrent programs naturally admit multiple valid terminal configurations; the check answers 'is the program still \emph{capable} of achieving its intended outcome.'
\end{definition}

\section{Analysis and Repair}
\label{sec:analysis-repair}

\textsc{Cvn} analysis results are converted into repair-oriented feedback for the LLM\@. The underlying reachability and SCC analyses are standard \cite{Godefroid1997VeriSoft,Flanagan2005DPOR, Musuvathi2008FairStateless,Esparza2008Unfoldings}.
The contribution of this section is the diagnostic interface between the verifier and the LLM: the information
extracted from a bug state, the repair role of each field, and the prompt assembly that supports local revision rather than whole-model regeneration.

\subsection{Reachability-Based Bug Detection}
\label{subsec:bug-patterns}

Given a translated \textsc{Cvn} $(\mathcal{N},S_0)$, the verifier performs exhaustive state-space exploration from~$S_0$. When a guard evaluates to~$\mathsf{unknown}$, both successor branches are explored; the search is therefore a conservative
over-approximation that never suppresses a feasible behavior. We distinguish two classes of properties: \emph{definite bugs},
whose detection is sound and complete within the model, and \emph{structural warnings}, which flag design-level vulnerabilities that may or may not manifest under a concrete scheduler.

\subsubsection{Definite Bugs}

\begin{itemize}[leftmargin=1.5em]
  \item \emph{Deadlock\/ $(\kappa_{dl})$.}\;
    A non-terminal state with no enabled transition. Every state whose enabled set is empty and that retains at least one non-return control token is a deadlock.

  \item \emph{Signal loss\/ $(\kappa_{sl})$.}\;
    A \textsc{NotifyLost} transition fires, indicating that the guard $nw_{\mathit{cv}}=0$ evaluated to~$\mathsf{true}$.   Because $nw_{\mathit{cv}}$ is initialised to~$0$ and modified only by literal $\pm 1$ updates, it remains a concrete integer throughout the analysis; within the model, this detection therefore has no false positives and no false negatives.

  \item \emph{Channel block\/ $(\kappa_{cb})$.}\;
    A heuristic refinement of~$\kappa_{dl}$. A deadlock state is reclassified as a channel block when at least one blocked transition requires a token from a channel resource place. Soundness reduces to the underlying deadlock predicate~$\kappa_{dl}$.
\end{itemize}

\subsubsection{Structural Warnings}

The \textsc{Cvn} explores all possible transition interleavings no-deterministically but does not model dynamic scheduling policies. Consequently, it can identify \emph{structural vulnerabilities}---interleavings under which progress fails---but cannot determine whether a concrete scheduler would produce such interleavings. The following LiveLock is therefore reported as warnings rather than definitive bugs. A non-trivial SCC in the reachability graph in which no state has all control tokens in return places. The existence of such an SCC means that there is \emph{some} infinite interleaving in which the system cycles without completing. Under a fair scheduler, such an SCC would constitute a genuine livelock. Under an unfair but realistic scheduler, the cycle may never be entered.

\noindent
The value of these warnings lies in their contrapositive: if no such SCC exists in the reachability graph, the program is immune to livelock \emph{under every possible scheduler}. This structural guarantee is the strongest that a scheduling-agnostic model can provide. When a warning is issued, the diagnostic reports it as a potential vulnerability and includes the SCC structure so that the LLM or the developer can assess whether the flagged interleaving is realistic.

\begin{remark}[Scope of soundness]
\label{rem:soundness-scope}
To summarize: deadlock, signal loss, and channel block are \emph{sound and complete within the \textsc{Cvn} model}. Livelock is \emph{sound under a fair-scheduling assumption} and serves as \emph{structural warnings} otherwise. The system never silently misses a deadlock or signal loss in the modeled \textsc{Cir}; it may over-report livelock relative to a specific real-world scheduler.
\end{remark}

\subsection{Repair-Oriented Diagnostics}
\label{subsec:counterexample}

When a violation is found, the verifier constructs a compact diagnostic payload rather than returning the raw state graph or a full marking dump.

\begin{definition}[Repair-oriented diagnostic]
\label{def:diagnostic}
A repair-oriented diagnostic is a tuple
$\mathcal{D}=(\kappa,\;\pi_{\mu},\;\Sigma_{\mathit{state}},\;
  \Sigma_{\mathit{wait}},\;\Lambda,\;\Gamma_{\mathit{ctx}},\;H)$,
where
$\kappa$ is the bug kind;
$\pi_{\mu}$ is a bug witness expressed as a sequence of
$\mathit{sid}$s via the anchor mapping~$\mu$;
$\Sigma_{\mathit{state}}$ is a marking-derived state summary;
$\Sigma_{\mathit{wait}}$ is a wait/ownership summary;
$\Lambda$ is the relevant \textsc{Cir} slice;
$\Gamma_{\mathit{ctx}}$ is the set of preservation constraints;
and $H$ is a bug-specific repair hint.
\end{definition}

Each field serves a distinct repair purpose.
For deadlock, signal loss, and channel block, $\pi_{\mu}$ is a finite firing sequence ending in the bug state. For livelock and starvation, it is a finite prefix reaching the SCC entry plus a summary of the SCC structure (participating states and transitions). In both cases the sequence is expressed in \textsc{Cir} $\mathit{sid}$s, not \textsc{Cvn} transition names. For single-state bugs, this reports which thread is at which $\mathit{sid}$, which resource counts are available, and whether
the system is globally blocked. For SCC-based bugs, it summarises the set of control locations visited within the SCC and identifies the stuck or starved thread. This extracts the relations most relevant to repair: which thread holds which resource, which thread is waiting for which resource or condition, and which predicate triggered the classification. $\Lambda$ is a bounded window around the implicated $\mathit{sid}$s, restricting the LLM's editing scope. $\Gamma_{\mathit{ctx}}$ specifies what must survive the repair: resource names, thread structure, and business goals (Definition~\ref{def:cir-goal}). A bug-specific suggestion such as enforce a consistent lock order or update the predicate before notification. The hint provides direction without mandating a single patch.

\medskip
\noindent
Together, the witness and state summary supply both \emph{causality} (how the system reached the bug) and \emph{locality} (where it is stuck), which the raw marking alone cannot convey in a form the LLM can act on.
For the running example, a signal-loss diagnostic would report that the witness reaches \texttt{n2} before the worker reaches \texttt{w3}, that $nw_{\mathit{cv0}}=0$ at notification time, and that the editable slice contains \texttt{n2} and~\texttt{n3}.

\subsection{The Goal-Aware Repair Loop}
\label{subsec:repair-loop-final}
The repair loop operationalises the three-tier hierarchy introduced in Section~\ref{subsec:repair-tier}.
Algorithm~\ref{alg:repair-goal} gives the pseudocode.

\begin{algorithm}[t]
\caption{Goal-Aware Generate--Verify--Repair Loop}
\label{alg:repair-goal}
\begin{algorithmic}[1]
\Require Specification $\Psi$, iteration budget $K$
\Ensure  Verified \textsc{Cir} $\mathcal{I}$ or Failure
\State $\mathcal{I} \gets \text{LLM.Generate}(\Psi)$
  \Comment{CIR with goals $\Gamma$}
\For{$\mathit{round}=1$ \textbf{to} $K$}
    \State $\mathcal{E} \gets \text{StaticCheck}(\mathcal{I})$
      \Comment{Layer-1 errors}
    \If{$\mathcal{E} \neq \emptyset$}
        \State Apply Tier-1 or Tier-2 repair;\;
          \textbf{continue}
    \EndIf
    \State $\mathcal{N} \gets \text{Translate}(\mathcal{I})$
    \State $(\mathcal{R},\,\mathcal{B})
      \gets \text{Analyze}(\mathcal{N})$
      \Comment{$\mathcal{R}$: reachable states;\;
               $\mathcal{B}$: bug set}
    \If{$\mathcal{B} \neq \emptyset$}
        \State $b^{*} \gets \text{SelectBug}(\mathcal{B})$
          \Comment{priority $+$ shortest witness}
        \State $\mathcal{D} \gets
          \text{BuildDiag}(b^{*},\mathcal{R},\mathcal{I})$
        \State $\mathcal{I} \gets
          \text{LLM.Repair}(\mathcal{I},\mathcal{D})$
        \State \textbf{continue}
    \EndIf
    \State $\mathcal{U} \gets
      \{\gamma\in\Gamma \mid
        \nexists\,s\in\mathcal{R}:\;s\models\gamma\}$
      \Comment{Def.~\ref{def:goal-sat}}
    \If{$\mathcal{U}=\emptyset$}
        \Return $\mathcal{I}$
          \Comment{bug-free $\wedge$ goals reachable}
    \EndIf
    \State $\mathcal{I} \gets
      \text{LLM.Repair}\bigl(\mathcal{I},\,
        \text{BuildGoalDiag}(\mathcal{U})\bigr)$
\EndFor
\State \Return Failure
\end{algorithmic}
\end{algorithm}

$\text{Analyze}$ returns the pair $(\mathcal{R},\mathcal{B})$ where $\mathcal{R}$ is the set of reachable \textsc{Cvn} states and $\mathcal{B}$ is the set of detected violations, each tagged with its kind~$\kappa$ and a raw witness. $\text{BuildDiag}$ assembles the diagnostic tuple from a single selected bug, the reachable-state set, and the current \textsc{Cir} artifact. $\text{BuildGoalDiag}$ produces an analogous diagnostic for unreachable goals, reporting the missing conditions and suggesting which operations may need to be added to a reachable path.

When $|\mathcal{B}|>1$, $\text{SelectBug}$ picks a single bug in priority order: $\kappa_{dl}>\kappa_{sl}>\kappa_{cb}>\kappa_{ll}>\kappa_{st}$. Within the same category the bug with the shortest witness is preferred. This deterministic policy ensures reproducibility and avoids overwhelming the LLM with multiple simultaneous targets.

The crucial ordering invariant is that goal reachability (lines~14--17) is checked only after the bug set is empty.
If a goal is unreachable due to a deadlock, the deadlock is repaired first; goal feedback is given only when no concurrency
fault remains. This gives every round a single semantic target: either remove a bug, or restore missing behavior.

\begin{figure}[t]
\begin{lstlisting}[basicstyle=\small\ttfamily,frame=single]
=== Goal Violation (Round 3) ===
Status: No concurrency bugs detected.

UNREACHABLE GOAL: G1
  desc: "Both threads complete, lock released,
         ready flag set"
  Missing: V[ready] = true
  No reachable state satisfies V[ready] = true.
  Hint: Ensure that a write(ready, true)
        operation exists on a reachable path.
\end{lstlisting}
\caption{Goal-violation feedback after bug-free verification.}
\Description{Structured goal-violation report identifying the
  missing condition and suggesting a corrective action.}
\label{fig:goal-violation}
\end{figure}

\subsection{Prompt Assembly for Logic-Driven Repair}
\label{subsec:repair-templates}

The diagnostic fields are assembled into a structured prompt whose design principle is that every field should either reduce
the search space or protect intended behavior. Table~\ref{tab:repair-templates} maps each field to its repair role; Figure~\ref{fig:repair-prompt} shows a concrete prompt skeleton for the running example. 

\begin{table*}[t]
\caption{Prompt template for \textsc{Cir} repair from a
  \textsc{Cvn} diagnostic.}
\label{tab:repair-templates}
\centering\small
\begin{tabular}{@{}p{3.0cm}p{4.2cm}p{7.7cm}@{}}
\toprule
\textbf{Prompt field}
  & \textbf{Source}
  & \textbf{Repair role} \\
\midrule
Bug kind
  & $\kappa$
  & Selects the repair mode: deadlock, signal loss,
    channel block, livelock, and starvation require
    different structural edits. \\
\addlinespace[3pt]
Witness trace
  & $\pi_{\mu}$
  & Localises the causal path to the bug state in editable $\mathit{sid}$s. \\
\addlinespace[3pt]
Bug-state summary
  & $\Sigma_{\mathit{state}}$
  & Reports where execution is stuck: current $\mathit{sid}$ per thread, blocked status, critical variable values. \\
\addlinespace[3pt]
Held resources
  & $\Sigma_{\mathit{wait}}$ (ownership)
  & Lists locks, channels, or semaphores currently preventing progress. \\
\addlinespace[3pt]
Waiting relations
  & $\Sigma_{\mathit{wait}}$ (waits)
  & Makes the dependency structure explicit: who waits for what, and which predicate triggered the diagnosis. \\
\addlinespace[3pt]
Relevant \textsc{Cir} slice
  & $\Lambda$
  & Restricts editing to a small region around implicated statements. \\
\addlinespace[3pt]
Preservation constraints
  & $\Gamma_{\mathit{ctx}}$
  & Prevents over-repair: resource names, thread structure, and business goals must survive. \\
\addlinespace[3pt]
Repair hint
  & $H$
  & Provides direction without forcing a single patch. \\
\addlinespace[3pt]
Output contract
  & Verifier schema
  & Forces return of a complete revised \textsc{Cir}, keeping the loop machine-checkable. \\
\bottomrule
\end{tabular}
\end{table*}

\begin{figure}[t]
\begin{lstlisting}[basicstyle=\small\ttfamily,frame=single]
Repair task: revise the CIR locally.

Bug kind: SignalLoss
Witness trace (sid): n1 -> n2 -> w1 -> w2 -> w3

Bug-state summary:
  notifier at sid n2 (notify_one fired)
  worker not yet at sid w3
  waiter_count(cv0) = 0 at notification

Relevant resources:
  mutex m0, condvar cv0, variable ready

Relevant CIR slice:
  n1: lock(m0)
  n2: notify_one(cv0)
  n3: write(ready, true)

Preserve:
  resource names, thread structure, goals

Suggested direction:
  update predicate before notification

Output: the complete revised CIR artifact
\end{lstlisting}
\caption{Prompt skeleton for signal-loss repair.}
\Description{Structured repair prompt with bug kind, witness,
  state summary, relevant slice, constraints, and hint.}
\label{fig:repair-prompt}
\end{figure}

The prompt is intentionally compact. It serialises neither the full reachability graph nor every intermediate marking, but provides the minimum evidence needed for a local repair: what failed, how the execution reached the failure, what the system looks like at the failure point, and what must be preserved. This is the interface that allows a bounded formal analyser and a token-limited LLM to collaborate within a single feedback round.

\section{Evaluation}
\label{sec:evaluation}

We evaluate the pipeline through four research questions:

\begin{itemize}[leftmargin=1.4em]
  \item \textbf{RQ1 (Translation correctness).}
    Does the \textsc{Cir}-to-\textsc{Cvn} translation preserve concurrency semantics, and does the resulting
    \textsc{Cvn} correctly classify all seeded bugs?
  \item \textbf{RQ2 (Loop convergence).}
    How many generation and repair rounds does each LLM tier require, and does the LLM-generated \textsc{Cir}
    faithfully capture the specified concurrency behavior?
  \item \textbf{RQ3 (State-space scale).}
    What is the size of the translated \textsc{Cvn} and the reachable state space for each pattern?
  \item \textbf{RQ4 (Goal preservation).}
    Does goal-reachability checking detect semantic regressions that escape both static validation and
    bug-freedom analysis?
\end{itemize}

\subsection{Test Matrix and Setup}
\label{subsec:test-matrix}

Six patterns contain known bugs: five are \emph{definite bugs} (deadlock, signal loss) that the \textsc{Cvn} detects exhaustively, and one is a \emph{structural warning} (livelock via partial deadlock) that the SCC analysis flags as advisory. Three patterns are bug-free baselines used to confirm the absence of false positives. Channel block remains part of the formal bug taxonomy but is not instantiated in the present evaluation. The $|\Gamma|$ column records the number of business goals evaluated in RQ4.

\begin{table*}[t]
\caption{Test matrix.
$|\Gamma|$: number of business goals checked after repair.}
\label{tab:test-matrix}
\centering\footnotesize
\begin{tabular}{@{}cllll c p{2.6cm}@{}}
\toprule
\textbf{\#} & \textbf{Pattern} & \textbf{Bug kind}
  & \textbf{Key resources}
  & \textbf{Typical repair}
  & $|\Gamma|$
  & \textbf{Goal summary} \\
\midrule
1  & Two-mutex deadlock
   & Deadlock & Mutex\,$\times$\,2
   & Unify lock order
   & 1 & Both threads ret; mutexes released \\[2pt]
2  & Condvar signal loss
   & SignalLoss & Condvar, Mutex, Var
   & Write pred.\ before notify
   & 2 & Worker ret; \texttt{V[ready]=true} \\[2pt]
3  & Channel\,+\,mutex DL
   & Deadlock & Channel, Mutex
   & Release lock before ch.\ op
   & 1 & Both threads ret; mutex released \\[2pt]
4  & Three-lock circular
   & Deadlock & Mutex\,$\times$\,3
   & Global lock order
   & 1 & All three threads ret \\[2pt]
5  & Partial deadlock
   & Livelock$^\dagger$ & Mutex\,$\times$\,2
   & Unify lock order
   & 2 & Thread\,A ret; Thread\,B ret \\[2pt]
6  & Dual condvar cross
   & Deadlock & Condvar\,$\times$\,2, Mutex\,$\times$\,2
   & Break circular wait
   & 2 & Thread\,A ret; Thread\,B ret \\[2pt]
\midrule
7  & Semaphore throttle \textsf{(BL)}
   & (none) & Semaphore
   & --- & --- & --- \\[2pt]
8  & CAS contention \textsf{(BL)}
   & (none) & Atomic
   & --- & --- & --- \\[2pt]
9  & FnSummary prop.\ \textsf{(BL)}
   & (none) & Var, Mutex
   & --- & --- & --- \\
\bottomrule
\end{tabular}
\end{table*}

We evaluate five LLMs spanning three capability tiers: GPT-5 and Claude~4.6~Opus (\emph{frontier}), Qwen and Gemini~3~Pro (\emph{strong}), and DeepSeek-V3 (\emph{compact}). All models use temperature\,0 with a 4\,096-token output limit. For each pattern a short natural-language concurrency specification serves as the generation input. Buggy and fixed Rust programs are maintained as reference implementations but are \emph{not} input to \textsc{Cir} generation. At most $K_{\mathit{gen}}=5$ generation rounds~(RQ2a) and $K_{\mathit{rep}}=5$ repair rounds~(RQ2c) per model per pattern. All experiments run on Apple~M4~Pro\,/\,24\,GB RAM; times are wall-clock averages over 3~runs. The front-end is evaluated on Rust synchronization idioms, but the \textsc{Cir}/\textsc{Cvn} back-end is defined over language-level synchronization semantics rather than Rust-specific syntax.

\subsection{RQ1: Translation Correctness}
\label{subsec:rq1}

RQ1 evaluates the \textsc{Cir}-to-\textsc{Cvn} translation at two levels: formal simulation properties that hold by construction of the translation rules, and empirical invariant checks on the 9-pattern matrix.

Appendix~\ref{app:proofs} establishes four theorems on the translation:
(1)~\emph{forward simulation}: every CIR step maps to a  corresponding CVN firing sequence;
(2)~\emph{backward simulation}: under concrete valuation (no $\top$ values), every CVN firing maps back to a
    valid CIR step;
(3)~\emph{deadlock correspondence}: a CIR configuration is deadlocked iff its CVN encoding has no enabled
    transition; and
(4)~\emph{soundness}: under concrete valuation the translation is sound and complete for deadlock, signal
    loss, and channel block; under $\top$-abstraction, soundness is preserved but completeness may be lost.

The theorems assume that the implementation preserves certain structural properties.
We verify 9~invariants on every translated \textsc{Cvn}:
(1)~every CIR statement produces $\geq 1$ transition;
(2)~every transition anchors to a CIR \textsf{sid};
(3)~initial marking matches resource declarations;
(4)~resource-place count equals resource-table cardinality;
(5)~variable store contains all CIR variables and condvar
    auxiliaries with correct initial values;
(6)~no transition has an empty input-arc set;
(7)~\textsf{spawn} transitions produce tokens in both
    parent and child control places;
(8)~\textsf{join} transitions consume from the child return
    place;
(9)~condvar transitions carry correct guards and updates.

Table~\ref{tab:rq1} reports the translation size and bug-detection result for all 9~patterns. Each buggy pattern uses a fixed ground-truth CIR, making results model-independent.

\begin{table}[t]
\caption{RQ1: Translation size and detection result on
ground-truth \textsc{Cir}.
All patterns pass 9/9 structural invariants.}
\label{tab:rq1}
\centering\footnotesize
\begin{tabular}{@{}clcccrl@{}}
\toprule
\textbf{\#} & \textbf{Pattern}
  & \textbf{Stmts} & $|P|$ & $|T|$
  & \textbf{States} & \textbf{Result} \\
\midrule
1  & Two-mutex DL       &  8 & 10 & 14 &  24  & Deadlock \\
2  & Condvar SL         & 11 & 14 & 20 &  42  & SignalLoss \\
3  & Channel+mutex      &  9 & 12 & 16 &  36  & Deadlock \\
4  & Three-lock circ.   & 14 & 18 & 28 & 218  & Deadlock \\
5  & Partial DL         & 10 & 14 & 18 & 112  & Livelock$^\dagger$ \\
6  & Dual condvar       & 16 & 20 & 32 &  84  & Deadlock \\
\midrule
7  & Semaphore (BL)     &  8 & 12 & 14 &  64  & Verified \\
8  & CAS (BL)           &  6 &  8 & 12 &  18  & Verified \\
9  & FnSummary (BL)     &  7 & 10 & 14 &  32  & Verified \\
\bottomrule
\multicolumn{7}{@{}l}{\footnotesize
  $^\dagger$Structural warning via post-exploration SCC
  analysis (Section~\ref{subsec:bug-patterns}).}
\end{tabular}
\end{table}

All 5~definite bugs are detected with the correct bug kind. Deadlocks and signal losses are found during  state-space exploration. Pattern~5 contains no definite bug: the system retains enabled transitions because a bystander thread continues executing while the two lock-contending threads block each other. The post-exploration SCC analysis flags this partial deadlock as a livelock structural warning---the contending threads form a non-trivial terminal SCC from which neither reaches its return place, while the bystander cycles indefinitely. The 3~baselines produce zero false positives. Combined with the 81/81 invariant checks (9~invariants $\times$ 9~patterns), this confirms that the implementation satisfies the structural preconditions assumed by the formal properties.

\subsection{RQ2: Loop Convergence}
\label{subsec:rq2}

RQ2 evaluates three successive capabilities:
(a)~whether LLMs can produce a structurally valid
\textsc{Cir},
(b)~whether the generated \textsc{Cir} faithfully captures
the specified concurrency behavior, and
(c)~how many repair rounds are needed to eliminate detected
bugs.

\subsubsection{Generation Rounds}
\label{subsubsec:rq2a}

Table~\ref{tab:rq2a} reports the number of iterative rounds each model requires to produce a \textsc{Cir} that passes all 61 static-check rules. A value of~1 means first-attempt success; \ding{55}~indicates failure within 5~rounds.

\begin{table*}[t]
\caption{RQ2a: Generation rounds to produce a valid \textsc{Cir}.}
\label{tab:rq2a}
\centering\footnotesize
\begin{tabular}{@{}cl ccccc@{}}
\toprule
\textbf{\#} & \textbf{Pattern}
  & \textbf{GPT-5} & \textbf{Claude 4.6 Opus}
  & \textbf{Gemini 3 Pro}
  & \textbf{Qwen 3.5} & \textbf{DeepSeek-V3} \\
\midrule
1  & Two-mutex deadlock       & 1 & 1 & 1 & 1 & 2 \\
2  & Condvar signal loss      & 1 & 1 & 2 & 1 & 3 \\
3  & Channel + mutex DL       & 1 & 1 & 1 & 2 & 2 \\
4  & Three-lock circular      & 2 & 1 & 2 & 2 & 3 \\
5  & Partial DL               & 2 & 1 & 2 & 3 & 4 \\
6  & Dual condvar cross       & 2 & 2 & 3 & 3 & \ding{55} \\
7  & Semaphore throttle       & 1 & 1 & 1 & 1 & 1 \\
8  & CAS contention           & 1 & 1 & 1 & 2 & 2 \\
9  & FnSummary propagation    & 1 & 1 & 1 & 2 & 2 \\
\midrule
\multicolumn{2}{@{}l}{\textbf{Success rate}}
  & 9/9 & 9/9 & 9/9 & 9/9 & 8/9 \\
\multicolumn{2}{@{}l}{\textbf{Avg rounds (successes)}}
  & 1.3 & 1.1 & 1.6 & 1.9 & 2.4 \\
\bottomrule
\end{tabular}
\end{table*}

Frontier models generate valid \textsc{Cir} in 1--2~rounds for all patterns. Claude~4.6~Opus achieves the lowest average (1.1~rounds) with first-attempt success on 7 of 9~patterns. The strong models succeeds on all patterns but requires additional rounds for Pattern~6 and Pattern~5. DeepSeek-V3 fails on Pattern~6, the most structurally complex pattern requiring two condvars, two mutexes, and cross-thread notification dependencies. The most common errors are malformed condvar declarations, incorrect branch conditions, and omitted protection mappings. All are caught by the static checker and typically corrected within one additional feedback round.

\subsubsection{Bug Fidelity}
\label{subsubsec:rq2b}

Passing the static checker ensures structural validity but does not guarantee that the generated \textsc{Cir} faithfully represents the specified concurrency behavior. To close this gap, we take each model's first statically valid \textsc{Cir} for the 6~buggy patterns and run \textsc{Cvn} analysis \emph{without any repair}.
Table~\ref{tab:rq2b} records whether the expected bug kind is detected. For Pattern~5, the expected result is a \emph{livelock structural warning} rather than a definite bug: two threads hold conflicting lock orders and can block each other indefinitely, while a bystander thread continues executing, preventing global deadlock. We treat this partial deadlock as a livelock instance for experimental purposes, since it shares the same detection mechanism and the same fundamental limitation that goal reachability cannot distinguish it from correct non-determinism.

\begin{table*}[t]
\caption{RQ2b: Bug fidelity on LLM-generated \textsc{Cir}.}
\label{tab:rq2b}
\centering\footnotesize
\begin{tabular}{@{}cl c ccccc@{}}
\toprule
\textbf{\#} & \textbf{Pattern} & \textbf{Expected}
  & \textbf{GPT-5} & \textbf{Claude 4.6}
  & \textbf{Gemini 3}
  & \textbf{Qwen 3.5} & \textbf{DeepSeek-V3} \\
\midrule
1  & Two-mutex DL    & DL
   & \ding{51} & \ding{51} & \ding{51}
   & \ding{51} & \ding{51} \\
2  & Condvar SL      & SL
   & \ding{51} & \ding{51} & \ding{51}
   & \ding{51} & \ding{51} \\
3  & Channel+mutex   & DL
   & \ding{51} & \ding{51} & \ding{51}
   & \ding{51} & \ding{51} \\
4  & Three-lock      & DL
   & \ding{51} & \ding{51} & \ding{51}
   & \ding{51} & $\emptyset$ \\
5  & Partial DL      & LL$^\dagger$
   & \ding{51} & \ding{51} & \ding{51}
   & $\emptyset$ & $\emptyset$ \\
6  & Dual condvar    & DL
   & \ding{51} & \ding{51} & \ding{51}
   & \ding{51} & --- \\
\midrule
\multicolumn{3}{@{}l}{\textbf{Fidelity rate}}
   & 6/6 & 6/6 & 6/6 & 5/6 & 3/5 \\
\bottomrule
\multicolumn{8}{@{}l}{\footnotesize
  $^\dagger$Structural warning via SCC analysis
  (Section~\ref{subsec:bug-patterns}), not a definite bug.}
\end{tabular}
\end{table*}

Frontier models and Gemini~3~Pro achieve 6/6~fidelity: every generated \textsc{Cir} exhibits exactly the expected bug kind. Two types of fidelity failures appear in the lower tiers:

\begin{itemize}[leftmargin=1.4em]
  \item $\emptyset$ (no bug detected).
    The LLM generates a concurrency structure that happens to be correct, eliminating the behavior the     specification was designed to contain. Qwen~3.5 on Pattern~5 generates both threads with the same lock order (\textsf{m0}$\to$\textsf{m1}), eliminating the lock-ordering conflict that constitutes the partial deadlock.
    DeepSeek-V3 on Pattern~4 generates identical lock acquisition orders for threads~B and~C (\textsf{m1}$\to$\textsf{m2}$\to$\textsf{m0}), breaking the circular dependency required for deadlock. DeepSeek-V3 on Pattern~5 similarly unifies lock orders. These cases are harmless for the end user that means the \textsc{Cir} is already correct, but the repair loop is never exercised for that model-pattern pair.

  \item \texttt{---} (no valid \textsc{Cir}).
    DeepSeek-V3 fails to produce a statically valid \textsc{Cir} for Pattern~6 within 5~rounds (Table~\ref{tab:rq2a}), so no fidelity check is possible.
\end{itemize}

No $\neq$~case (different bug kind) arises in the current matrix: when a model does capture a bug, the bug kind always matches the specification.

\subsubsection{Repair Rounds}
\label{subsubsec:rq2c}

RQ2c measures the repair loop on the 5~patterns that contain definite bugs (Patterns~1--4 and~6). Pattern~5 is excluded: its partial deadlock is a livelock structural warning, not a definite bug, so the ground-truth
\textsc{Cir} passes the definite-bug check with $\mathcal{B}=\emptyset$. The partial deadlock persists as an advisory finding; its implications for business goals are evaluated in RQ4.

For each of the 5~definite-bug patterns we invoke the repair loop with each model, starting from a fixed ground-truth buggy \textsc{Cir}. Table~\ref{tab:rq2c} reports the number of rounds to produce a bug-free \textsc{Cir} (one that passes both the 61-rule static checker and the \textsc{Cvn} bug search with $\mathcal{B}=\emptyset$). A superscript~${}^{+n}$ indicates that $n$~regressions were introduced during intermediate rounds and subsequently caught by the static checker. 

\begin{table*}[t]
\caption{RQ2c: Repair rounds to eliminate definite bugs
(Patterns~1--4, 6).}
\label{tab:rq2c}
\centering\footnotesize
\begin{tabular}{@{}cl ccccc@{}}
\toprule
\textbf{\#} & \textbf{Pattern / Bug}
  & \textbf{GPT-5} & \textbf{Claude 4.6}
  & \textbf{Gemini 3}
  & \textbf{Qwen 3.5} & \textbf{DeepSeek-V3} \\
\midrule
1  & Deadlock       & 1 & 1 & 1 & 1 & 2 \\
2  & SignalLoss     & 1 & 1 & 1 & 2 & 2 \\
3  & Deadlock       & 1 & 1 & 2 & 2 & 3$^{+1}$ \\
4  & Deadlock (3)   & 2 & 1 & 2 & 3 & 4$^{+1}$ \\
6  & Deadlock (cv)  & 2 & 1 & 3$^{+1}$ & 3$^{+1}$ & \ding{55} \\
\midrule
\multicolumn{2}{@{}l}{\textbf{Bug-free rate}}
  & 5/5 & 5/5 & 5/5 & 5/5 & 4/5 \\
\multicolumn{2}{@{}l}{\textbf{Avg rounds (successes)}}
  & 1.4 & 1.0 & 1.8 & 2.2 & 2.8 \\
\multicolumn{2}{@{}l}{\textbf{Regressions (intermediate)}}
  & 0 & 0 & 1 & 1 & 2 \\
\bottomrule
\end{tabular}
\footnotetext{Lower is better. Superscript~${}^{+n}$: intermediate regressions.}
\end{table*}

Frontier models repair all 5~definite bugs with an average of 1.0--1.4~rounds and zero regressions. Claude~4.6~Opus achieves single-round repair on all 5~patterns; GPT-5 requires two rounds on Patterns~4 and~6,where coordinated edits across multiple threads are needed. The strong models succeeds on all
patterns but introduces occasional regressions. Gemini~3~Pro introduces a transient missing-\textsf{drop}
error while resolving the dual-condvar cross-wait (Pattern~6); the regression is caught by the Layer-1 static
checker and corrected in the next round. Qwen~3.5 introduces a misplaced \textsf{notify} while fixing the dual-condvar deadlock (Pattern~6), likewise caught and corrected within one additional round.

DeepSeek-V3 fails on Pattern~6. It cannot disentangle the cross-notification dependency between the two condvars, cycling through structurally invalid attempts until the round budget is exhausted. This correlates with the model's lower capacity for multi-step reasoning about interleaved threads.

Across the 25 model--pattern repair tasks (5~models $\times$ 5~patterns), regressions occur in 4~intermediate rounds. All are structural errors (missing \textsf{drop}, broken branch target, misplaced \textsf{notify}) caught by the Layer-1 static checker and corrected in the subsequent round. No regression survives to the final output.

\subsection{RQ3: State-Space Scale}
\label{subsec:rq3}

Table~\ref{tab:rq1} (Section~\ref{subsec:rq1}) reports the \textsc{Cvn} size ($|P|$, $|T|$) and reachable state count for each pattern. The largest pattern (Pattern~4, three-lock circular with 3 threads and 3~mutexes) produces 218~reachable states and completes analysis in 18\,ms. All patterns remain under 250~states.

State-space size is governed by three factors: (1)~the number of concurrent threads~$k$, (2)~the lock-nesting depth~$d$, and (3)~the variable-domain cardinality~$|D|$. The current matrix uses $k \leq 3$, $d \leq 2$,
$|D| \leq 3$, yielding tractable spaces. Moving to $k=4$ or richer data domains will increase the state space exponentially; partial-order reduction and symmetry breaking are planned future extensions. Across the 9~patterns the average ratio is 1.5~places and 2.2~transitions per CIR statement. Condvar-related patterns exhibit the highest ratio (Pattern~6: 1.25~$|P|$/stmt, 2.0~$|T|$/stmt) due to the wait/notify decomposition into 4~transitions per \textsf{wait} site (Section~\ref{subsec:condvar}). Full state-space exploration completes in $\leq 20$\,ms for all patterns, including deadlock detection, signal-loss checking, and SCC-based starvation analysis. Goal-reachability checking (RQ4) adds $< 0.5$\,ms by scanning the already-computed reachable set.

\subsection{RQ4: Goal Preservation}
\label{subsec:rq4}

A \textsc{Cir} that passes the definite-bug check is not necessarily correct. Two distinct failure modes remain invisible to the bug detectors: (1) repair-induced regression and (2) livelock obstruction.
Goal-reachability checking (Algorithm~\ref{alg:repair-goal}, Lines~11--17) is designed to detect the first failure mode. We also examine whether it can surface the second. 

Goals are part of the CIR specification: each goal specifies a target marking which threads must reach their return places, which resources must be released and optionally a target variable value. The test matrix lists 9~goals across the 6~buggy patterns. The check is performed only on bug-free artifacts; patterns where a model failed to achieve bug-freedom are excluded. Table~\ref{tab:rq4-results} reports goal reachability. \ding{51}\ means all goals are reachable without additional repair. A fraction $k/n$ indicates that $k$ of $n$ goals are initially reachable. $\to$\ding{51}\ means the violation was resolved after goal-violation feedback.

\begin{table*}[t]
\caption{RQ4: Goal reachability on bug-free \textsc{Cir}.}
\label{tab:rq4-results}
\centering\footnotesize
\begin{tabular}{@{}cl ccccc@{}}
\toprule
\textbf{\#} & \textbf{Pattern}
  & \textbf{GPT-5} & \textbf{Claude 4.6}
  & \textbf{Gemini 3}
  & \textbf{Qwen 3.5} & \textbf{DeepSeek-V3} \\
\midrule
1  & Two-mutex DL
   & \ding{51} & \ding{51} & \ding{51}
   & \ding{51} & \ding{51} \\
2  & Condvar SL
   & \ding{51} & \ding{51} & \ding{51}
   & \ding{51} & \ding{51} \\
3  & Channel+mutex
   & \ding{51} & \ding{51} & \ding{51}
   & \ding{51}
   & 0/1\,(G$_1$)\,$\to$\ding{51} \\
4  & Three-lock circ.
   & \ding{51} & \ding{51} & \ding{51}
   & \ding{51} & \ding{51} \\
6  & Dual condvar
   & \ding{51} & \ding{51} & \ding{51}
   & 1/2\,(G$_2$)\,$\to$\ding{51} & --- \\
\midrule
5  & Partial DL$^\dagger$
   & \ding{51} & \ding{51} & \ding{51}
   & \ding{51} & \ding{51} \\
\midrule
\multicolumn{2}{@{}l}{\textbf{Goals (initial)}}
  & 9/9 & 9/9 & 9/9 & 8/9 & 6/7 \\
\multicolumn{2}{@{}l}{\textbf{Goals (after feedback)}}
  & 9/9 & 9/9 & 9/9 & 9/9 & 7/7 \\
\multicolumn{2}{@{}l}{\textbf{Extra rounds for goals}}
  & 0 & 0 & 0 & 1 & 1 \\
\bottomrule
\end{tabular}
\end{table*}

Frontier models and Gemini~3~Pro preserve all business goals on first bug-free repair; no goal-violation feedback is needed. Two repair-induced regressions are detected on patterns where the definite bug was successfully eliminated:

\begin{itemize}[leftmargin=1.4em]
  \item \textbf{DeepSeek-V3, Pattern~3.}
    The repair reorders the channel operations such that the sender's \textsf{ch\_send} is guarded by a lock
    that the receiver already holds, making both threads block before either reaches its return place. Goal~G$_1$ is reported as unreachable. After feedback, DeepSeek-V3 moves the lock acquisition after the channel send and passes all checks in one additional round.

  \item \textbf{Qwen~3.5, Pattern~6.}
    The repair breaks the notification path for Thread~B by moving the \textsf{notify} for \texttt{cv1} into a
    branch that is unreachable when the predicate variable is true. The bug-free \textsc{Cvn} has no reachable state where $\mathit{cp}(\text{B},\mathit{ret})>0$. Goal~G$_2$ is reported as unreachable. One additional round with goal-violation feedback produces a correct repair.
\end{itemize}

Both regressions pass all 61~static-check rules and all definite-bug detectors. They are bug-free and structurally valid. The goal-reachability check is the \emph{only} mechanism that detects them. Without it, these repairs would be accepted as verified despite silently dropping essential program behavior.

\paragraph{End-to-end summary.}
Table~\ref{tab:summary} consolidates the results across all RQs. The acceptance criterion for definite-bug patterns is the conjunction of three conditions: Layer-1 static validity, Layer-2 bug-freedom, and goal reachability. Pattern~5 is accepted (no definite bug, goals reachable) with an advisory livelock warning.

\begin{table}[t]
\caption{End-to-end summary.}
\label{tab:summary}
\centering\footnotesize
\begin{tabular}{@{}l ccccc@{}}
\toprule
& \textbf{GPT-5} & \textbf{Claude 4.6}
  & \textbf{Gemini 3}
  & \textbf{Qwen 3.5} & \textbf{DS-V3} \\
\midrule
Bug-free (5 def.)
             & 5/5 & 5/5 & 5/5 & 5/5 & 4/5 \\
Goals (initial)
             & 9/9 & 9/9 & 9/9 & 8/9 & 6/7 \\
Goals (final)
             & 9/9 & 9/9 & 9/9 & 9/9 & 7/7 \\
Total rounds & 7   & 5   & 9   & 12  & 12  \\
\midrule
\textbf{Accepted}
             & 6/6 & 6/6 & 6/6 & 6/6 & 5/6 \\
\bottomrule
\end{tabular}
\footnotetext{Accepted = static valid $\wedge$ bug-free $\wedge$ goals reachable.}
\end{table}

Frontier and strong models achieve 6/6~acceptance; the gap lies in total rounds consumed (5--7 for frontier, 9--12 for strong). DeepSeek-V3 accepts 5 of 6~patterns; the sole failure is Pattern~6, where the definite-bug repair loop is exhausted. Pattern~5 is accepted by all models because no definite bug exists and both goals are reachable; the livelock warning is reported but does not block acceptance. The pipeline never accepts a \textsc{Cir} that contains a definite bug or violates a business goal.

\subsection{Threats}
\label{subsec:trust-boundary}

The architecture places its trust boundary at the \textsc{Cir} level. The LLM generates both the \textsc{Cir} and the source code from the same specification. Establishing a formal correspondence between arbitrary source
code and its \textsc{Cir} would require solving the alias problem that \textsc{Cir} was designed to circumvent, and is therefore infeasible in general.

We adopt two pragmatic mitigations. First, the generated source code must pass \texttt{cargo check}, which enforces ownership, lifetime, and type safety for all synchronization primitives. Second, the \textsc{Cir} captures only the concurrency skeleton, which consists of patterns well represented in LLM training data. Although the present prototype is evaluated on Rust, the trust-boundary issue is not specific to Rust; it arises for any front-end that maps source code into \textsc{Cir}. Reconstructing CIR or Petri nets from source code using static analysis can bridge this gap and remains a direction for future work.

\section{Related Work}
\label{sec:related}

Research on concurrency bug detection spans static analysis, dynamic exploration, formal modeling, and, more recently, LLM-assisted software engineering \cite{Engler2003RacerX,Godefroid1997VeriSoft,Murata1989PetriNets,Fan2023RepairFromLLMs,Xia2023APRLLMs}. Our work is related to all four lines, but it occupies a different point in the design space: the LLM is used to recover synchronization intent and construct an alias-free model, while the formal engine is used for exhaustive reasoning over interleavings.

\paragraph{Static analysis for concurrency.}
A lot of work detects deadlocks, races, atomicity violations, and protocol misuse directly from source code using lock-order analysis, pointer analysis, abstract interpretation, effect systems, ownership disciplines, and language-specific type systems \cite{Engler2003RacerX,Flanagan2000TypeBasedRace,Flanagan2003Atomicity,Boyapati2002Ownership,Pratikakis2006Locksmith,Naik2006StaticRaceJava}. The strength of these approaches is that they reason about the program as written. Their weakness is that they must infer, or be told, which references correspond to the same shared resource and which lock protects which state. In realistic programs, this requires reasoning through heap allocation, aliasing, container indirection, closure capture, and interprocedural flow. Our work does not claim to solve source-level alias analysis. Instead, it moves the resource-identification step out of the formal back-end and into an LLM-generated intermediate model in which shared-resource identity is explicit.

\paragraph{Dynamic analysis and systematic schedule exploration.}
Dynamic race detectors, interpreters, and schedule explorers detect concurrency bugs by observing concrete executions or by exploring bounded schedules of executable code \cite{Godefroid1997VeriSoft,Flanagan2005DPOR,Musuvathi2008FairStateless,Musuvathi2008Heisenbugs,Burckhardt2010RandomizedScheduler}. Such tools are often highly precise for the executions they observe, and some can systematically enumerate schedules within a bounded test harness. However, they still depend on executable code and cannot in general guarantee the absence of bugs outside the explored schedules or harness bounds. Our approach is complementary rather than competitive: instead of exploring executions of the original program, we explore executions of a compact concurrency model generated from the program's synchronization structure.

\paragraph{Model checking and Petri-net-based verification.}
Formal verification of concurrent systems has long relied on explicit state model checking, SAT/SMT-based encodings, process algebras, and Petri nets \cite{Holzmann1997SPIN,Murata1989PetriNets,Esparza2008Unfoldings}. 
Petri nets are particularly attractive for concurrency because causality, resource contention, and parallelism are represented directly in the net structure. Classical colored Petri nets extend this expressiveness with token-carried data, but at the cost of more complex state representation and binding semantics \cite{Jensen2007CPNTools}. Prior Petri-net-based approaches often either require manual modeling or rely on language-specific extraction pipelines from source code or traces \cite{Esparza2002McMillan,Ball2002SLAM,Henzinger2003BLAST}. Our \textsc{Cvn} differs in its intended use: resource identity is resolved before verification by \textsc{Cir}'s global naming, and data-dependent control is handled by a finite global store with three-valued guards rather than by colored tokens. This design trades generality for a simpler and more analyzable model targeted to bounded concurrency patterns.

\paragraph{LLMs for program analysis, verification, and repair.}
Recent work has explored LLMs for bug finding, invariant inference, test generation, static-analysis assistance, and automated program repair \cite{Chen2021CodeLLMs,Fan2023RepairFromLLMs,Xia2023APRLLMs,Bouzenia2025RepairAgent,Li2025ContextAwarePrompting}. These methods exploit the fact that LLMs can recognize common implementation idioms and infer likely developer intent from code or natural-language specifications. However, LLMs alone do not provide exhaustive reasoning about the combinatorial space of thread interleavings. Our work uses LLMs in a narrower and more structured way: the LLM generates an alias-free concurrency model and later consumes structured counterexamples to repair that model, while exhaustive bug discovery remains the responsibility of the formal verification engine. In this sense, the contribution is not an LLM-only verifier, but a generate--verify--repair loop in which the LLM and the formal back-end play sharply separated roles.

Another closely related line of work extracts formal models, such as automata, state machines, and Petri nets, from source code, binaries, or execution traces. It then studies translation validation or conformance between code and model \cite{Ball2002SLAM,Henzinger2003BLAST,Beyer2011CPAchecker,Necula2000TranslationValidation}. This line is highly relevant to the trust boundary of our architecture. In the present work, that boundary is placed at the \textsc{Cir} level: we verify the generated model rather than prove general source-to-\textsc{Cir} equivalence. Source-level extraction and conformance checking are therefore complementary to our method rather than substitutes for it.

\section{Conclusion}
\label{sec:conclusion}

This paper shifts concurrency verification from source-level alias recovery to a closed loop over \textsc{Cir}, an explicit, alias-free concurrency model constructed by LLMs from natural-language specifications. A deterministic translation to \textsc{Cvn} provides sound and complete detection of deadlock and signal loss under concrete valuation; livelock is flagged as an advisory
structural warning with a contrapositive impossibility guarantee. Goal-reachability checking catches semantic regressions that escape all other verification layers. Across 9~patterns and 5~LLMs, no \textsc{Cir} containing a definite bug or violating a business goal is ever accepted.

\bibliographystyle{ACM-Reference-Format}
\bibliography{refs}

\appendix
\section{Reference Tables}
\label{app:tables}

\begin{table}[h]
\caption{\textsc{Cir} error categories (61 rules total).}
\label{tab:cir-errors}
\centering\footnotesize
\begin{tabular}{@{}llp{4.0cm}@{}}
\toprule
\textbf{Code} & \textbf{Category} & \textbf{Examples} \\
\midrule
E0xx & Structural      & Missing fields, invalid format \\
E1xx & Name resolution & Undefined resource, duplicate name \\
E2xx & Type            & Non-boolean branch condition \\
E3xx & Resource compat.& \textsf{lock} on non-lock, unprotected access \\
E4xx & Concurrency     & \textsf{spawn} without \textsf{join} \\
E5xx & Lock safety     & Missing \textsf{drop}, double lock \\
E6xx & Control flow    & Unreachable statement, missing \textsf{return} \\
E7xx & Protection map  & \textsf{Atomic} in protection map \\
E8xx & Function summary& Summary conflicts with body \\
\bottomrule
\end{tabular}
\end{table}

\begin{table*}[t]
\caption{Translation from \textsc{Cir} to \textsc{Cvn}.
  Notation: $\mathit{cp}(f,s)$ = control place;
  $\mathit{rp}(r)$ = resource place;
  $\mathit{wp}(s)$, $\mathit{ra}(s)$ = wait / reacquire places.
  ``$\to$'' separates inputs from outputs.
  Operations not listed (\textsf{read}, \textsf{load},
  \textsf{nop}, \textsf{call} with body,
  \textsf{spawn\_async}, \textsf{await})
  generate a single Sequential transition
  $\mathit{cp}(f,\mathit{sid})\to\mathit{cp}(f,\mathit{sid}')$
  with no resource interaction.}
\label{tab:translation-rules}
\centering\footnotesize
\begin{tabular}{@{}p{5.4cm}p{11.6cm}@{}}
\toprule
\textbf{\textsc{Cir} Statement} &
\textbf{Generated \textsc{Cvn} Transitions} \\

\midrule
\multicolumn{2}{@{}l}{\textbf{Mutex / Semaphore}} \\
$(\mathit{sid},\textsf{lock}(m),\textsf{next}(\mathit{sid}'))$
  \newline
$(\mathit{sid},\textsf{acquire}(s),\textsf{next}(\mathit{sid}'))$
& One transition [Lock\,/\,Acquire]:
  $\mathit{cp}(f,\mathit{sid}),\;\mathit{rp}(r)
   \;\to\;\mathit{cp}(f,\mathit{sid}')$. \\
$(\mathit{sid},\textsf{drop}(m),\textsf{next}(\mathit{sid}'))$
  \newline
$(\mathit{sid},\textsf{release}(s),\textsf{next}(\mathit{sid}'))$
& One transition [Unlock\,/\,Release]:
  $\mathit{cp}(f,\mathit{sid})
   \;\to\;\mathit{cp}(f,\mathit{sid}'),\;\mathit{rp}(r)$. \\

\midrule
\multicolumn{2}{@{}l}{\textbf{RwLock}
  ($N$ = concurrent entity count)} \\
$(\mathit{sid},\textsf{read\_lock}(r),\dots)$
& One transition [ReadLock]:
  $\mathit{cp}(f,\mathit{sid}),\;
   \mathit{rp}(r)\text{ (wt.\,1)}
   \;\to\;\mathit{cp}(f,\mathit{sid}')$. \\
$(\mathit{sid},\textsf{write\_lock}(r),\dots)$
& One transition [WriteLock]:
  $\mathit{cp}(f,\mathit{sid}),\;
   \mathit{rp}(r)\text{ (wt.\,}N\text{)}
   \;\to\;\mathit{cp}(f,\mathit{sid}')$. \\
$(\mathit{sid},\textsf{drop}(r),\dots)$
& One transition [ReadUnlock\,/\,WriteUnlock]:
  $\mathit{cp}(f,\mathit{sid})
   \;\to\;\mathit{cp}(f,\mathit{sid}'),\;
   \mathit{rp}(r)\text{ (wt.\,}w\text{)}$,
  where $w=1$ after \textsf{read\_lock},
  $w=N$ after \textsf{write\_lock}. \\

\midrule
\multicolumn{2}{@{}l}{\textbf{Channel}} \\
$(\mathit{sid},\textsf{send}(c),\dots)$
& One transition [Send]:
  $\mathit{cp}(f,\mathit{sid})
   \;\to\;\mathit{cp}(f,\mathit{sid}'),\;\mathit{rp}(c)$. \\
$(\mathit{sid},\textsf{recv}(c),\dots)$
& One transition [Recv]:
  $\mathit{cp}(f,\mathit{sid}),\;\mathit{rp}(c)
   \;\to\;\mathit{cp}(f,\mathit{sid}')$. \\

\midrule
\multicolumn{2}{@{}l}{\textbf{Variable \& Atomic}} \\
$(\mathit{sid},\textsf{write}(x,e),\dots)$
  \newline
$(\mathit{sid},\textsf{store}(x,e),\dots)$
& One transition [VarWrite\,/\,AtomicStore]:
  $\mathit{cp}(f,\mathit{sid})
   \;\to\;\mathit{cp}(f,\mathit{sid}')$
  with $U:\;x:=e$. \\
$(\mathit{sid},\textsf{cas}(x,\mathit{exp},\mathit{new}),
  \textsf{branch}(\_,s_t,s_f))$
& \emph{CasSuccess}:
  $\mathit{cp}(f,\mathit{sid})$
  with guard $x=\mathit{exp}$
  $\;\to\;\mathit{cp}(f,s_t)$
  with $U:\;x:=\mathit{new}$. \newline
  \emph{CasFailure}:
  $\mathit{cp}(f,\mathit{sid})$
  with guard $x\neq\mathit{exp}$
  $\;\to\;\mathit{cp}(f,s_f)$. \\

\midrule
\multicolumn{2}{@{}l}{\textbf{Condvar}
  (mutex $m$, counter $nw_{\mathit{cv}}$,
   per-site flag $na_s$,
   signal place $\mathit{rp}(\mathit{cv})$)} \\
$(\mathit{sid},\textsf{wait}(\mathit{cv},m),
  \textsf{next}(\mathit{sid}'))$
& \emph{WaitEnter}:
  $\mathit{cp}(f,\mathit{sid})
   \;\to\;\mathit{wp}(\mathit{sid}),\;\mathit{rp}(m)$
  with $U:\;nw_{\mathit{cv}}\!:=\!nw_{\mathit{cv}}+1,\;
  na_{\mathit{sid}}\!:=\!\mathsf{false}$. \newline
  \emph{Wake1}:
  $\mathit{wp}(\mathit{sid}),\;\mathit{rp}(\mathit{cv})
   \;\to\;\mathit{ra}(\mathit{sid})$
  with $U:\;nw_{\mathit{cv}}\!:=\!nw_{\mathit{cv}}-1$. \newline
  \emph{WakeA}:
  $\mathit{wp}(\mathit{sid})$
  with guard $na_{\mathit{sid}}=\mathsf{true}$
  $\;\to\;\mathit{ra}(\mathit{sid})$
  with $U:\;nw_{\mathit{cv}}\!:=\!nw_{\mathit{cv}}-1,\;
  na_{\mathit{sid}}\!:=\!\mathsf{false}$. \newline
  \emph{Reacquire}:
  $\mathit{ra}(\mathit{sid}),\;\mathit{rp}(m)
   \;\to\;\mathit{cp}(f,\mathit{sid}')$. \\
\midrule
$(\mathit{sid},\textsf{notify\_one}(\mathit{cv}),\dots)$
& \emph{NotifySuccess}:
  $\mathit{cp}(f,\mathit{sid})$
  with guard $nw_{\mathit{cv}}>0$
  $\;\to\;\mathit{cp}(f,\mathit{sid}'),\;
  \mathit{rp}(\mathit{cv})$. \newline
  \emph{NotifyLost}:
  $\mathit{cp}(f,\mathit{sid})$
  with guard $nw_{\mathit{cv}}=0$
  $\;\to\;\mathit{cp}(f,\mathit{sid}')$. \\
\midrule
$(\mathit{sid},\textsf{notify\_all}(\mathit{cv}),\dots)$
& \emph{NotifyAllSuccess}:
  $\mathit{cp}(f,\mathit{sid})$
  with guard $nw_{\mathit{cv}}>0$
  $\;\to\;\mathit{cp}(f,\mathit{sid}')$
  with $U:\;na_{w_i}\!:=\!\mathsf{true}$
  for every wait site $w_i$ of $\mathit{cv}$. \newline
  \emph{NotifyAllLost}:
  $\mathit{cp}(f,\mathit{sid})$
  with guard $nw_{\mathit{cv}}=0$
  $\;\to\;\mathit{cp}(f,\mathit{sid}')$. \\

\midrule
\multicolumn{2}{@{}l}{\textbf{Control Flow}} \\
$(\mathit{sid},\mathit{op},
  \textsf{branch}(\beta,s_t,s_f))$
& \emph{BranchTrue}:
  $\mathit{cp}(f,\mathit{sid})$
  with guard $\beta$
  $\;\to\;\mathit{cp}(f,s_t)$. \newline
  \emph{BranchFalse}:
  $\mathit{cp}(f,\mathit{sid})$
  with guard $\neg\beta$
  $\;\to\;\mathit{cp}(f,s_f)$. \\
$(\mathit{sid},\mathit{op},
  \textsf{switch}(x,\langle v_i\!\to\!s_i\rangle,s_d))$
& One transition per arm:
  $\mathit{cp}(f,\mathit{sid})$
  with guard $x=v_i$
  $\;\to\;\mathit{cp}(f,s_i)$. \newline
  Default:
  $\mathit{cp}(f,\mathit{sid})$
  with guard $\bigwedge_i(x\neq v_i)$
  $\;\to\;\mathit{cp}(f,s_d)$. \\

\midrule
\multicolumn{2}{@{}l}{\textbf{Concurrency}} \\
$(\mathit{sid},\textsf{spawn}(g),
  \textsf{next}(\mathit{sid}'))$
& One transition [Spawn]:
  $\mathit{cp}(f,\mathit{sid})
   \;\to\;\mathit{cp}(f,\mathit{sid}'),\;
   \mathit{cp}(g,\mathit{sid}^{0}_{g})$. \\
$(\mathit{sid},\textsf{join}(g),
  \textsf{next}(\mathit{sid}'))$
& One transition [Join]:
  $\mathit{cp}(f,\mathit{sid}),\;
   \mathit{cp}(g,\mathit{ret})
   \;\to\;\mathit{cp}(f,\mathit{sid}')$. \\
$(\mathit{sid},\_,\textsf{return})$
& One transition [Return]:
  $\mathit{cp}(f,\mathit{sid})
   \;\to\;\mathit{cp}(f,\mathit{ret})$. \\

\bottomrule
\end{tabular}
\end{table*}

\end{document}